\documentclass[11pt]{article}

\usepackage{geometry}
\usepackage{graphicx}
\usepackage{amsmath, amssymb, amsfonts, amsthm, float}  
\usepackage{enumerate, color, framed, float, multirow}
\usepackage{comment, longtable, caption, subcaption, appendix}
\usepackage[sort,longnamesfirst]{natbib}
\usepackage{setspace, parskip}
\usepackage{placeins}

\allowdisplaybreaks

\newcommand{\pcite}[1]{\citeauthor{#1}'s \citeyearpar{#1}}

\newcommand{\ds}{\displaystyle}
\newcommand{\E}{\text{E}}
\newcommand{\Var}{\text{Var}}
\newcommand{\X}{\mathsf{X}}

\newcommand{\B}{\mathcal{B}}
\newcommand{\real}{{\mathbb R}}

\newcommand\numberthis{\addtocounter{equation}{1}\tag{\theequation}}

\theoremstyle{remark}

\newtheorem{remark}{Remark}

\newtheorem{example}{Example}

\begin{document}
\title{Revisiting the Gelman-Rubin Diagnostic}
\date{\today}
\author{
Dootika Vats\thanks{Research funded by the National Science Foundation}\\
Department of Mathematics and Statistics\\
Indian Institute of Technology Kanpur\\
Kanpur, IN - 208016\\
\texttt{dootika@iitk.ac.in}
\and
Christina Knudson\thanks{Research funded by the University of St. Thomas Office of Study Abroad,  College of Arts and Science, and Center for Faculty Development} \\
Department of Mathematics\\
University of St. Thomas\\
St. Paul, Minnesota 55105\\
\texttt{knud8583@stthomas.edu}
}
\maketitle

\begin{abstract}
\pcite{gelm:rubi:1992a} convergence diagnostic  is one of the most popular methods for terminating a Markov chain Monte Carlo (MCMC) sampler. Since the seminal paper, researchers have developed sophisticated methods for estimating variance of Monte Carlo averages. We show that these estimators find immediate use in the Gelman-Rubin statistic, a connection not previously established in the literature. We incorporate these estimators to upgrade both the univariate and multivariate Gelman-Rubin statistics, leading to improved stability in MCMC termination time. An immediate advantage is that our new Gelman-Rubin statistic can be calculated for a single chain. In addition, we establish a one-to-one relationship between the Gelman-Rubin statistic and  effective sample size. Leveraging this relationship, we develop a principled termination criterion for the Gelman-Rubin statistic. Finally, we demonstrate the utility of our improved diagnostic via examples. 
\end{abstract}

\section{Introduction} % (fold)
\label{sec:introduction}

In the early 1990s, a surge in  Markov chain Monte Carlo (MCMC) research produced a variety of convergence diagnostics, including those developed by  \cite{gewe:1992},  \cite{gelm:rubi:1992a}, and \cite{raft:lewi:how:1992}.
 %In the early 1990s, research in Markov chain Monte Carlo (MCMC) diagnostics surged,  producing convergence diagnostics developed by \cite{gewe:1992},  \cite{gelm:rubi:1992a}, and \cite{raft:lewi:how:1992}; a detailed discussion can be found in \cite{cowl:carl:1996}.  
%researchers prolifically produced diagnostics for  Markov chain Monte Carlo (MCMC)  convergence. This included the diagnostics of \cite{gewe:1992},  \cite{raft:lewi:how:1992}, and  \cite{gelm:rubi:1992a}. 
The Gelman-Rubin (GR) diagnostic has been one of the most popular diagnostics for MCMC convergence:  Google Scholar indicates the original paper has been cited over 9000 times, with over 1000  citations in 2017 alone.  Primary reasons for its popularity are its ease of use and its widespread availability in software.

%and (ii) do we have enough samples to approximate the target distribution. The first convergence deals with the popular idea of ``burn-in.'' The GR diagnostic addresses the second question, which is the focus of this manuscript. 

The GR diagnostic framework relies on $m$ parallel MCMC chains, each run for $n$ steps with starting points determined by a distribution that is over-dispersed  relative to the target distribution. 
% Convergence is diagnosed when the empirical distribution obtained from $mn$ samples is close to the target distribution.
% More specifically, t
The GR statistic (denoted $\hat{R}$) is the square root of the ratio of two estimators for the target variance.  In finite samples, the numerator overestimates this variance and the denominator underestimates it. Each estimator converges to the target variance, meaning that $\hat{R}$ converges to 1 as $n$ increases. When $\hat{R}$ is sufficiently close to 1, the GR diagnostic declares convergence. 

%More specifically, the GR diagnostic (denoted $\hat{R}$) is the square root of the ratio of two asymptotically unbiased estimators of the target variance. However, in finite samples, the numerator overestimates this variance and the denominator underestimates it, resulting in $\hat{R} > 1$ for finite samples and generally decreases. When $\hat{R}$ sufficiently nears one, the GR diagnostic declares convergence. 

\cite{gel:car:2004} recommend terminating simulation when $\hat{R} \leq 1.1$. This threshold has been adopted widely by practitioners. 
 Table~\ref{tab:GR_literature} summarizes  the $\hat{R}$ thresholds reported by 100 randomly sampled papers that cited \cite{gelm:rubi:1992a} in 2017.  The recommended cutoff of 1.1 was used by 43 of the 100. The next most commonly used cutoffs were 1.01 and 1.05.  A cutoff higher than 1.1  was used by 10 papers,  and the smallest threshold was 1.003. 
%We randomly sampled 100 papers that cited \cite{gelm:rubi:1992a} in 2017 and noted the cutoff used. Papers that did not report a criterion were excluded. Table~\ref{tab:GR_literature} shows a summary of the $\hat{R}$ thresholds reported. Of the 100 papers, 43 used the recommended 1.1 cutoff and 10 others used a higher cutoff; 1.01 and 1.05 were the other two most common cutoffs, with the smallest threshold being 1.003. 
\begin{table}[h]
	\caption{Distribution of $\hat{R}$ cutoff used over 100 sampled papers.}
	\label{tab:GR_literature}
	\centering
	\begin{tabular}{|l|c|c|c|c|c|c|c|c|c|c|c|}
	\hline
$\hat{R}$ cutoff & 1.003 & 1.01 & 1.02 & 1.03 & 1.04 & 1.05 & 1.06 & 1.07 &  1.1 &  1.2 &  1.3 \\ \hline
 Count&   1  &  12   &  9  &   9  &   2   & 11    & 2   &  1   & 43  &   9   &  1 \\ \hline
	\end{tabular}
\end{table}

\begin{figure}[tb]
	\centering
	\includegraphics[width = 3in]{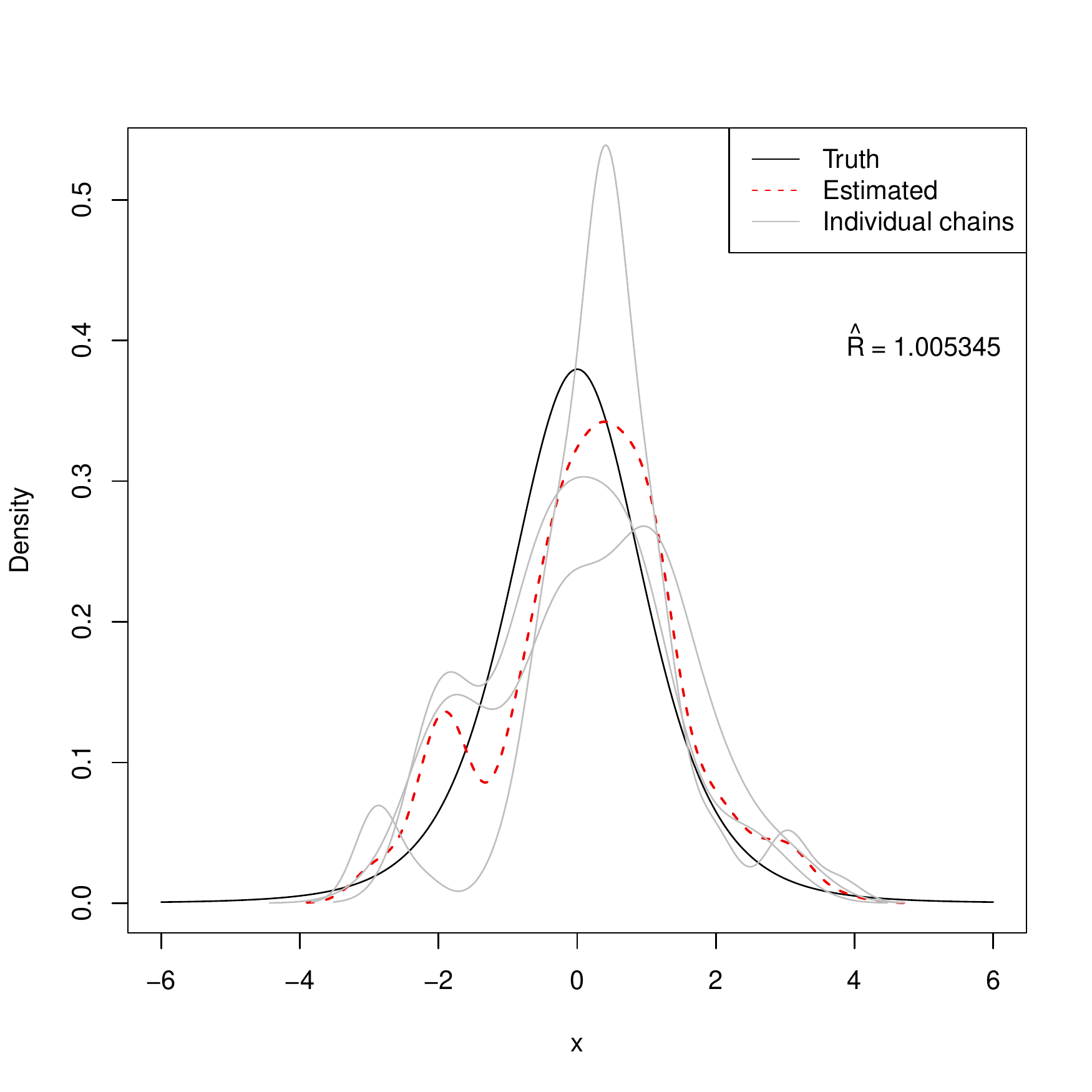}
	\caption{Density estimate and $\hat{R}$ along with the three individual density estimates.}
	\label{fig:t_density}
\end{figure}

We argue that a cutoff of $\hat{R} \leq 1.1$ is much too high to yield reasonable estimates of target quantities. Consider the example of sampling from a $t_5$-distribution, a $t$ distribution with 5 degrees of freedom, using a random walk Metropolis-Hastings sampler with a $N(\, \cdot \,, 2.6^2)$ proposal. We run $m = 3$ chains for $2n = 150$ steps, with starting values drawn from a $t_2$ distribution. We discard the first $n$ samples from each chain, as recommended by \cite{gelm:rubi:1992a}. Density estimates from the three chains are in Figure~\ref{fig:t_density}. The resulting $\hat{R}$ from this run is $1.0053$, which is much smaller than the termination threshold suggested by  \cite{gelm:rubi:1992a}, but the estimated density is far from the truth. 

The termination threshold critically impacts the quality of estimation, yet current practices do not suffice. The suggested threshold of 1.1 seems arbitrary and---as the example suggests---may be much too high to yield confidence in final estimates. We respond with two contributions: (i) we present an improved GR statistic and (ii) we establish a principled method of selecting a GR diagnostic termination threshold. 

First, we propose improving both the univariate GR statistic (Section~\ref{sec:univariate_diagnostic}) and the multivariate GR statistic of \cite{bro:gel:1998} (Section~\ref{sec:multivariate_potential_scale_reduction_factor}) 
 by using recently-developed estimators of the variance of Monte Carlo averages. The relative efficiency of the original estimator used in the GR statistic versus our new estimator grows without bound as the chain length increases, resulting in dramatic stabilization of the GR statistic. 
Such an improvement was indirectly {implied}  by \cite{fleg:hara:jone:2008}. 

Second, in Section~\ref{sec:choosing_delta}, we present a method of selecting a principled, interpretable GR statistic cutoff by identifying a one-to-one correspondence between the GR statistic and the effective sample size (ESS) for estimating the mean of the target distribution.
 Specifically, we show that
\[ 
\hat{R} \approx \sqrt{1 + \dfrac{m}{\text{ESS}}}\,.
\]
Thus, for $m$ chains, choosing a termination threshold $1.1$ implies an ESS of approximately $5m$, or five independent samples per chain; this is clearly too low to estimate the mean with any reasonable certainty.

In Section~\ref{sec:example}, we assess the performance of our methods in four examples. First, we present a complete analysis of the $t_5$-distribution example. 
The second example uses an autoregressive model---for which underlying true variances are known---to assess the two statistics'  time-to-convergence stability as it compares to the truth. The third example  compares the traditional and updated GR statistics' performance for a bimodal target distribution. We consider two cases, the first when the Markov chain gets stuck in a local mode, and the second when the Markov chain is able to jump between modes. The fourth and final example demonstrates the implementation of our improved GR statistic on a Bayesian logistic regression model analyzing the Titanic dataset; this highlights the marked improvement in the regression estimates' stability when using the ESS-based termination threshold. We end with a discussion in Section~\ref{sec:discussion}.

\section{Markov chains and convergence} % (fold)
\label{sec:markov_chain_and_convergence}

Let $F$ be a target distribution defined on a space $\X$ equipped with a countably generated $\sigma$-field, $\B(\X)$.
Let $P:\X \times \B(\X) \to [0,1]$ denote a Markov chain transition kernel such that for $x \in \X$ and $A \in \B(\X)$, $P(x,A) = \Pr(X_2 \in A |X_1 = x)$. For $i = 1, \dots, m$, let $\{X_{it}\}_{t \geq 1}$ denote the $i$th independent Markov chain. The starting value of the Markov chains, $X_{i1}$, are user-chosen and are either fixed or drawn randomly from a convenient initial distribution. We assume that $P$ is $F$-invariant and Harris ergodic \cite[see][for definitions]{meyn:twee:2009} for definitions so that $P$ converges to $F$ (in total variation distance) for any initial distribution. 

Typically, MCMC returns samples that are correlated and only approximately from $F$. This has allowed for diverse literature on the issue of convergence of an MCMC algorithm. There are two main types of convergence that are relevant to most MCMC problems \cite[see][for a detailed discussion]{roy:2019,vats:rob:fle:jon:2020}: (i) the convergence of the $n$-step Markov transition, $P^n$, to the stationary distribution $F$, and (ii) the convergence of sample statistics to the truth.  The first is often termed as the ``burn-in'' problem, where a first chunk of the samples is discarded when the starting distribution of $X_{i1}$ is far away from $F$. Determining how many samples to retain is a challenging problem that often involves a detailed study of the specific Markov chain kernel, $P$. See \cite{rose:1995a,jone:hobe:2001} for a theoretical exposition. 

Convergence diagnostics that address the second class of convergence either assess convergence of the empirical distribution function or assess convergence of moments of functions of interest. Many density-based diagnostics have been proposed in the literature. \cite{boone:merr:2014} measure the Hellinger distance between estimated marginal densities from multiple chains. A similar approach was used by \cite{hjor:vade:2005} with a distance metric similar to the Kullback Leibler (KL) divergence and by \cite{dix:roy:2017} with a KL divergence and adaptive kernel density estimators. \cite{van:schmid:2017} use a state-space partition of the clusters in the MCMC output to diagnose convergence of the estimated target distribution.

This article and the original Gelman-Rubin diagnostic focuses on diagnosing moment-based convergence. That is, if interest is in estimating the mean, quantile, variance, etc of $F$, the MCMC process is said to have converged when the sample statistics are close enough to the truth. Convergence is  guaranteed due to Harris ergodicity of the chains. Suppose interest is in estimating the mean of the posterior distribution $F$, $\mu = \E_FX_{it}$. The Monte Carlo average of each chain estimates $\mu$ consistently. That is, as $n\to \infty$, due to the Markov chain strong law
\[
\bar{X}_{i \cdot} = \dfrac{1}{n} \ds \sum_{t=1}^{n} X_{it} \overset{a.s.}{\to} \mu\,.
\]
The combined estimator of $\mu$ from the $m$ Markov chains is
\[
\hat{\mu} = \dfrac{1}{m}\sum_{i=1}^{m} \bar{X}_{i\cdot} \overset{a.s.}{\to} \mu\,,
\]
as $n\to \infty$. If $m = 1$, then $\hat{\mu} = \bar{X}_{1\cdot}$. Moment-based diagnostic tools measure the quality of estimation of $\hat{\mu}$. \cite{gewe:1992} constructed a hypothesis test for testing the equality of the means of two nonoverlapping sections of a Markov chain. These sections are usually the first 10\% and the last 50\% of the chain. Similarly, \cite{raft:lewi:how:1992} used a two-state Markov chain assumption to construct a univariate diagnostic based on estimating quantiles of univariate components of the target distribution. A comprehensive survey of these and other diagnostics can be found in \cite{cowl:carl:1996} which also extends the above to the case when interest is in estimating the mean of a function $g:\X \to \mathbb{R}^p$. Almost all of the methods require the estimation of the limiting variance $\tau^2_{\infty}: = \lim_{n\to \infty} n \text{Var}(\bar{X}_{i\cdot})$, which is finite if a Markov chain central limit theorem holds \cite[see][for conditions]{jone:2004}. That is, a Markov chain central limit theorem holds if there exists $\tau^2_{\infty}< \infty$ such that as $n \to \infty$
\begin{equation}
\label{eq:clt}
\sqrt{n}(\bar{X}_{i \cdot} - \mu) \overset{d}{\to} N(0, \tau^2_{\infty})\,.
\end{equation}
\cite{fleg:gong:2015,gong:fleg:2016,jone:hara:caff:neat:2006} propose a family of sequential termination rules that stop simulation the first time the variability in $\hat{\mu}$ is (relatively) small.  For the sequential stopping rules to yield confidence regions with the nominal coverage probability, estimators of $\tau^2_{\infty}$ must be strongly consistent, a property that has been shown for a wide range of estimators, including batch means \citep{liu:fleg:2018,vats:fleg:jones:2019}, spectral variance \citep{fleg:jone:2010,vats:fleg:jon:2018} and regeneration-based estimators \citep{jone:hara:caff:neat:2006,seil:1982}.   

By far, the most popular method for terminating an MCMC sampler run is the Gelman-Rubin diagnostic of \cite{gelm:rubi:1992a} and \cite{bro:gel:1998}. In the following sections, we introduce the Gelman-Rubin diagnostic in detail and reformulate the univariate and multivariate diagnostics facilitating the use of strongly consistent estimators of $\tau^2_{\infty}$. This allows us to find a novel connection between ESS and the Gelman-Rubin statistic, thus motivating a termination threshold for the Gelman-Rubin statistic. As in \cite{gelm:rubi:1992a}, we assume throughout that a Markov chain central limit theorem holds.

% section markov_chain_and_convergence (end)

% section motivating_example (end)
\section{Univariate diagnostic} % (fold)
\label{sec:univariate_diagnostic}
% Univariate and multivariate diagnostics will lead to fundamentally different convergence criterions and hence are studied separately.

% In this section we arrive at the potential scale reduction factor for an MCMC run of $m$ independent chains, where $m$ is also allowed to be 1.

\subsection{Original Gelman-Rubin statistic} % (fold)
\label{sub:original_gr}
Let $F$ be the target distribution with mean $\mu \in \mathbb{R}$ and variance $\sigma^2 < \infty$. 
% Let $X_{it}$ denote the Markov chain draw from chain $i$ ($i = 1, \dots, m$) at time $t$ ($t = 1, \dots, n$). That is, the $i$th Markov chain sequence is $X_{i1}, \dots, X_{in}$. 
\cite{gelm:rubi:1992a} construct two estimators of $\sigma^2$ and compare the square root of their ratio to 1. This process is described below. 

Recall that $\bar{X}_{i \cdot}$ is the sample mean from chain $i$ and $\hat{\mu}$ is the overall mean. Let $s_i^2$ denote the sample variance for chain $i$ and $s^2$ be the average of the $m$ sample variances. That is,
\begin{equation*}
\label{eq:s_i}
	s_i^2 = \dfrac{1}{n-1} \sum_{t=1}^{n} (X_{it} - \bar{X}_{i \cdot})^2 \qquad \text{ and } \qquad s^2 = \dfrac{1}{m} \sum_{i=1}^{m} s_i^2\,.
\end{equation*}
Although $s^2$ is strongly consistent for $\sigma^2$ as $n \to \infty$, it is biased for $\sigma^2$ for non-trivial Markov chains.   In fact,
\begin{equation}
\label{eq:sigma2_expectation}
\E_F(s_i^2) = \dfrac{n}{n-1} \left(\sigma^2 - \Var_F(\bar{X}_{i\cdot})  \right)\,.
\end{equation}
% \begin{align*}
% \dfrac{1}{n}\sum_{t=1}^{n} (X_{it} - \mu)^2 &= \dfrac{1}{n}\sum_{t=1}^{n} (X_{it} - \bar{X}_{i \cdot})^2 + (\bar{X}_{i \cdot} - \mu)^2
% \intertext{Taking expectation,}
% \sigma^2 &= \dfrac{n-1}{n} \E(s_i^2) + \Var{(\bar{X}_{i\cdot})}\numberthis \label{eq:sigma2_expectation}\\
% \Rightarrow \E(s_i^2) &= \dfrac{n}{n-1} \left(\sigma^2 - \Var(\bar{X}_{i\cdot})  \right)\,.
% \end{align*}
When the samples are independent and identically distributed, $\Var_F(\bar{X}_{i\cdot}) = \sigma^2/n$ and $s_i^2$ is an unbiased estimator of $\sigma^2$. However, for samples obtained through MCMC, $\Var_F(\bar{X}_{i\cdot})$ is often much larger than $\sigma^2/n$ due to positive correlation in the Markov chain. Thus, $s_i^2$ on average, underestimates the target variance. 

Define $\tau^2_n:= n\Var_F(\bar{X}_{i\cdot})$ and let $\tau^2_\infty = \lim_{n \to \infty} \tau^2_n < \infty$.  \cite{gelm:rubi:1992a} perform a bias correction by estimating $\tau^2_n/n = \Var_F(\bar{X}_{i \cdot})$ with $B$, the sample variance of sample means from $m$ chains. That is, 
\begin{equation}
\label{eq:gr_B}
\dfrac{B}{n} = \dfrac{1}{m-1} \ds \sum_{i=1}^{m} (\bar{X}_{i \cdot} - \hat{\mu} )^2\,.	
\end{equation}
Using  \eqref{eq:gr_B} to estimate $\Var_F(\bar{X}_{i \cdot})$ in \eqref{eq:sigma2_expectation} yields the following  estimator of $\sigma^2$:
\begin{equation*}
\label{eq:gr_unbiased_sigma2}
	\hat{\sigma}^2:= \dfrac{n-1}{n} s^2 + \dfrac{B}{n}\,. 
\end{equation*}
The univariate GR potential scale reduction factor (PSRF) is 
\begin{equation}
\label{eq:gr_psrf}
 \hat{R} = \sqrt{\dfrac{\hat{\sigma}^2}{s^2}}\,. 
\end{equation}
\cite{gelm:rubi:1992a} and \cite{gel:car:2004} argue that if an over-dispersed starting distribution for the Markov chain is used, $\hat{\sigma}^2$ overestimates $\sigma^2$---due to \eqref{eq:sigma2_expectation}---and $s^2$ underestimates $\sigma^2$. Since both are consistent for $\sigma^2$, $\hat{R}$ decreases to 1 as $n$ increases. Simulation is stopped when $\hat{R} \leq  \delta$ for some $\delta > 1$.

\begin{remark}\label{rem:difference_origin}
The  univariate PSRF presented by \cite{bro:gel:1998} and  \cite{gelm:rubi:1992a}  differs from \eqref{eq:gr_psrf}. Specifically, it is defined as
\[\sqrt{\hat{R}} = \sqrt{\dfrac{\hat{\sigma}^2 +  \frac{B}{mn}}{s^2} \dfrac{df+3}{df+1}}\,, \]
where $df$ is the degrees of freedom for the numerator estimated via a method of moments. Since this original estimator, $\hat{R}$ has evolved into the expression in \eqref{eq:gr_psrf}. Popular resources for MCMC convergence diagnostics such as \cite{gel:car:2004} and softwares such as \texttt{Stan} use the expression in \eqref{eq:gr_psrf}. Although the \texttt{R} package \texttt{coda} uses the original GR expression, we commit to the expression in \eqref{eq:gr_psrf}.
\end{remark}

\subsection{New univariate PSRF} % (fold)
\label{sec:improved_estimation_of_g_r_diagnostic}

% Recall that  $\hat{\sigma}^2$ overestimates $\sigma^2$ for an overdispersed starting distribution and is unbiased if the starting distribution is the stationary distribution. However, the variability in $\hat{\sigma}^2$ is considerably increased by the inefficient estimation of $\Var(\bar{X}_{i\cdot})$. 
Our improved construction of the PSRF incorporates efficient estimators of $\Var_F(\bar{X}_{i\cdot})$. Due to the correlation in the Markov chain,
\[\tau^2_n = n\Var_F\left( \bar{X}_{i \cdot}  \right)  =  \Var_F(X_{i1}) + 2 \sum_{k=1}^{n-1} \left( \dfrac{n-k}{n}  \right)\text{Cov}_F\left(X_{i1}, X_{i(1+k)} \right)\,.\]
% As $n \to \infty$, $\tau^2_n \to \tau^2_{\infty}$, which is the asymptotic variance in the Markov chain CLT (see for example \cite{jone:2004}). 
We use known estimators of $\tau^2_{\infty}$ to estimate $\tau^2_n$; in fact these estimators are technically estimating $\tau^2_n$ but are consistent for $\tau^2_{\infty}$ as $n\to \infty$. A significant amount of research in the past two decades  has resulted in improved estimation of $\tau^2_{\infty}$. This includes batch means estimators  and regenerative estimators \citep{jone:hara:caff:neat:2006}, spectral variance estimators and overlapping batch means estimators \citep{fleg:jone:2010}, and weighted batch means estimators \citep{liu:fleg:2018}. Under appropriate conditions, the estimators above are strongly consistent but are biased from below for $\tau^2_{\infty}$ \citep{vats:flegal:2018}. The initial sequence estimators of \cite{gey:1992} are asymptotically conservative but only apply  to reversible Markov chains. 

We use a lugsail version of the replicated batch means estimator to estimate $\tau^2_n$. As \cite{vats:flegal:2018} describe, the lugsail estimator is biased from above in finite samples but strongly and mean square consistent for $\tau^2_\infty$ as $n \to \infty$. Thus, even without an over-dispersed starting distribution, the lugsail estimator yields a biased-from-above estimate of $\tau^2_n$.  In order to combine variance estimates from multiple chains, we use a replicated version of the lugsail batch means estimator \`a la \cite{arg:and:2006}. This replicated estimator accounts for the case when independent copies of the chain are concentrated in different areas of the support of the distribution, a case that arises often in multi-modal targets. We now describe the replicated lugsail batch means estimator.

Suppose $n$ is such that $n = a\cdot b$ where $a$ is the number of batches and $b$ is the batch size. Both $a$ and  $b$ must increase with $n$; usual choices of $b$ include $ \lfloor n^{1/3} \rfloor $ and $ \lfloor n^{1/2} \rfloor$.  For the $i$th chain, define the mean for batch $k = 1, \dots, a$ as
\[
\bar{Y}_{ik} = \dfrac{1}{b} \sum_{t = (k-1)b + 1}^{kb}X_{it}\,. 
\]
The replicated batch means estimator of $\tau^2_n$ is,
\begin{equation*}
\label{eq:tau2_i_bm}
\hat{\tau}_{b}^2 := \dfrac{b}{am-1}	\ds \sum_{i=1}^{m} \sum_{k=1}^{a} (\bar{Y}_{ik} - \hat{\mu})^2\,.
\end{equation*}
Here the subscript $b$  in $\hat{\tau}_{b}^2$ indicates the batch size used to construct the estimator. The replicated lugsail batch means estimator is,
\begin{equation}
\label{eq:lugsail_i}
	\hat{\tau}^2_{L} := 2\hat{\tau}_{b}^2 - \hat{\tau}_{b/3}^2 \,,
\end{equation}
where $\hat{\tau}_{b/3}^2$ is the replicated batch means estimator constructed using batch size $\lfloor b/3 \rfloor$. 
% The final estimator of $\tau^2_n$ is the average of the $m$ variance estimators,
% \[\hat{\tau}^2_L := \dfrac{1}{m} \sum_{i=1}^{m} \hat{\tau}^2_{iL}\,. \]

An advantage of $\hat{\tau}^2_{L}$ over $B$ is its relative efficiency. 
The large sample variance of $B$ is $2 \tau^4_{\infty} /(m-1)$ \citep{gelm:rubi:1992a} while
 the large-sample variance of the replicated lugsail batch means estimator is $6 \tau^4_{\infty}/am$ \citep{gupta:vats:2020,arg:and:2006}. Because $a$ increases with $n$, the large sample relative efficiency of $B$ versus {$\hat{\tau}^2_L$} is
\begin{equation}
	\label{eq:efficiency_comparison}
	\text{eff}(B, \hat{\tau}^2_L) = \dfrac{m a}{3(m-1)} \to \infty \qquad \text{ as } n \to \infty\,. 
\end{equation}
Equation \eqref{eq:efficiency_comparison} shows that as the Markov chain length increases, the relative variance of $B$ versus $\hat{\tau}^2_L$ grows; for any reasonable choice of $n$, the replicated lugsail batch means estimator is markedly more efficient than $B$. Section~\ref{sec:example} will show that incorporating $\hat{\tau}^2_L$ rather than $B$ dramatically stabilizes the termination of MCMC.

Using $\hat{\tau}^2_L$ instead of $B$ yields the following biased-from-above estimator of $\sigma^2$
\begin{equation*}
\label{eq:VK_sigma2}
	 \hat{\sigma}^2_{L} := \dfrac{n-1}{n}s^2 + \dfrac{\hat{\tau}_L^2}{n}\,. 
\end{equation*}
Using $\hat{\sigma}^2_L$ instead of $\hat{\sigma}^2$ in $\hat{R}$ yields the following improved estimator for the PSRF:
\begin{equation}
\label{eq:vk_psrf}
	\hat{R}_L = \sqrt{\dfrac{\hat{\sigma}^2_L}{s^2}}\,.
\end{equation}
As before, the criterion for terminating simulation is $\hat{R}_L \leq  \delta$ for some $\delta > 1$.

\section{Multivariate PSRF} % (fold)
\label{sec:multivariate_potential_scale_reduction_factor}
\subsection{Original multivariate PSRF} % (fold)
\label{sub:original_mpsrf}

Most MCMC problems are inherently multivariate in that the goal is to sample from a multidimensional target distribution. Acknowledging the multivariate nature of estimation is critical in order to account for the interdependence between components of the chain \citep{vats:fleg:jones:2019}.
%As \cite{vats:fleg:jones:2019} conclude, it is critical to acknowledge this multivariate nature of estimation so as to account for the interdependence between components. 
\cite{bro:gel:1998} proposed the following multivariate extension of the univariate GR diagnostic. 

Let $F$ be a $p$-dimensional target distribution with mean $\mathbf{\mu} \in \real^p$ and let $\Sigma$ be the $p \times p$ variance-covariance matrix of the target distribution. Let $\mathbf{X}_{i1}, \dots, \mathbf{X}_{in}$ be the $i$th parallel Markov chain; each $\mathbf{X}_{it} = (X_{it1}, \dots, X_{itp})^T \in \real^p$. Let $\bar{\mathbf{X}}_{i\cdot} = n^{-1} \sum_{t=1}^{n} \mathbf{X}_{it}$ be the mean vector of the $i$th chain and let the overall mean be $\hat{\mathbf{\mu}} = m^{-1} \sum_{i=1}^{m} \bar{\mathbf{X}}_{i\cdot}$. Let $S_i$ be the sample covariance matrix for chain $i$, and let $S$ be the sample mean of  $S_1, \ldots, S_m$. That is
\[
S_i = \dfrac{1}{n-1} \ds \sum_{t=1}^{n} (\mathbf{X}_{it} - \bar{\mathbf{X}}_{i\cdot})(\mathbf{X}_{it} - \bar{\mathbf{X}}_{i\cdot})^T \qquad \text{ and } \qquad S = \dfrac{1}{m} \sum_{i=1}^{m} S_i\,.
\]
Just as in the univariate case, \cite{bro:gel:1998}  decompose the target variance:
\[\Sigma = \left(\dfrac{n-1}{n}\right)\E_F(S) + \Var_F(\bar{\mathbf{X}}_{i\cdot}) \,.  \]
Let $T_{n} := n\Var_F(\mathbf{\bar{X}}_{i \cdot})$ and let $T_{\infty} = \lim_{n \to \infty}n\Var_F(\mathbf{\bar{X}}_{i \cdot})$. Then $T_{\infty}$ is the asymptotic variance-covariance matrix in the multivariate Markov chain CLT. When $p=1$, $T_n = \tau^2_n$ and $T_{\infty} = \tau^2_{\infty}$. 
\cite{bro:gel:1998} estimate $T_n/n$ with the sample covariance matrix of the sample mean vectors from $m$ chains.  Define $\mathbf{B}$ such that
\[
\dfrac{\mathbf{B}}{n} = \dfrac{1}{m-1} \sum_{i = 1}^{m} (\mathbf{\bar{X}}_{i\cdot} - \mathbf{\hat{\mu}})(\mathbf{\bar{X}}_{i\cdot} - \mathbf{\hat{\mu}})^T\,. 
\]
Using $\mathbf{B}/n$ to correct for the bias in $S$ yields 
\[\widehat{\Sigma} := \left( \dfrac{n-1}{n}\right)S +  \dfrac{\mathbf{B}}{n}\,.\]
As in the univariate case, the goal is to compare the ratio of these estimators of $\Sigma$. However, because $\Sigma$ is a $p \times p$ matrix, a univariate quantification of this ratio is required. Let $\lambda_{\max}(A)$ denote the largest eigenvalue of a matrix $A$. The multivariate PSRF is
\begin{equation}
\label{eq:original_mpsrf}
	\hat{R}^p = \sqrt{\left( \dfrac{n-1}{n} \right) + \dfrac{\lambda_{\max}(S^{-1} \mathbf{B})}{n} }\,.
\end{equation}

\begin{remark}\label{rem:difference_original}
As in the univariate case, we use a different expression from the original paper by \cite{bro:gel:1998} so that the multivariate expression here is a direct generalization of the univariate PSRF. The expression in \cite{bro:gel:1998} is
\[ 
	\hat{R}^p = \sqrt{\left( \dfrac{n-1}{n} \right) + \left(\dfrac{m+1}{m} \right)\dfrac{\lambda_{\max}(S^{-1} \mathbf{B})}{n} }\,.
\]
% {\color{red} but this is the version that is used in most software now, right? If so (I hope so), I think we should mention this so it doesn't seem like we are doing our own thing just because it's convenient.}
\end{remark}

The estimator $\mathbf{B}$ will not be positive definite in the realistic event of $m$ being smaller than $p$. Also, the use of the largest eigenvalue is likely the reason the multivariate PSRF has not found large practical use in the literature. The largest eigenvalue quantifies the variability in the direction of the largest variation, the principal eigenvector of $S^{-1}\mathbf{B}$. This can be significantly larger than any of the individual variances, thus leading to a needlessly conservative termination criterion.

% So our understanding is that using the largest eigenvalue of $S^{-1}\mathbf{B}$ ensured that $\hat{R}^p$ is positive. This use of the largest eigenvalue is likely the reason the multivariate PSRF has not been used as much as the univariate PSRF. The largest eigenvalue quantifies the variability in the direction of the largest variation. This can be significantly larger than any of the individual variances. 

\subsection{New multivariate PSRF} % (fold)
\label{sub:new_multivariate_psrf}

% {\color{red} can we be more specific about what we are using from each of the cited papers that follow? Also I'm confused about who provide the positive def estimators (Kosorok, liu and flegal, and vats and flegal?) and who's providing the recent work (just dai and jones?) due to the commas. I can rearrange the sentence if you tell me what's what!}
Recent work by \cite{dai:jon:2017}, \cite{koso:2000}, \cite{liu:fleg:2018}, \cite{vats:flegal:2018}, and \cite{vats:fleg:jon:2018} provide estimators of $T_{\infty}$. We use the biased-from-above, multivariate replicated lugsail batch means estimator to estimate $T_n$. As before, for the $i$th chain, define the mean vector for batch $k = 1, \dots, a$ as
\[
\bar{\mathbf{Y}}_{ik} = \dfrac{1}{b} \sum_{t = (k-1)b + 1}^{kb}\mathbf{X}_{it}\,. 
\]
The multivariate replicated batch means estimator of $T_n$ is,
\begin{equation*}
\label{eq:tau2_i_bm}
\hat{T}_{b} := \dfrac{b}{am-1}	\ds \sum_{i=1}^{m} \sum_{k=1}^{a} (\bar{\mathbf{Y}}_{ik} - \hat{\mu})(\bar{\mathbf{Y}}_{ik} - \hat{\mu})^T\,.
\end{equation*}
The multivariate replicated lugsail batch means estimator for the $i$th chain is,
\begin{equation}
\label{eq:lugsail_i_multi}
	\hat{T}_{L} := 2\hat{T}_{b} - \hat{T}_{b/3} \,.
\end{equation}
% where $\hat{\tau}_{b/3}$ is the replicated batch means estimator constructed using batch size $b/3$. 
% The final estimator of $\tau^2_n$ is the average of the $m$ variance estimators,
% \[\hat{\tau}^2_L := \dfrac{1}{m} \sum_{i=1}^{m} \hat{\tau}^2_{iL}\,. \]
%
% Let $\hat{T}_{L}$ be the natural multivariate equivalent of the replicated lugsail batch means in \eqref{eq:lugsail_i}.
 % lugsail batch means estimator of $T_n$ from chain $i$; this is a direct multivariate generalization of the univariate lugsail batch means estimators. Let $\hat{T}_L$ be the averaged matrix from the $m$ chains, that is,
% \[\hat{T}_L = \dfrac{1}{m} \ds \sum_{i=1}^{m} \hat{T}_{iL}\,.\]
% 
Define 
\[
\hat{\Sigma}_L =  \left( \dfrac{n-1}{n}  \right) S + \dfrac{\hat{T}_L}{n}\,.
\]
Let $\det(\cdot)$ denote determinant. We define our multivariate PSRF as 
\begin{equation}
\label{eq:vk_msprf}
	\hat{R}^p_{L} = \sqrt{\left( \dfrac{n-1}{n} \right) + \dfrac{\det( S^{-1} \hat{T}_L)^{1/p}}{n} }\,.
\end{equation}
\begin{remark}\label{rem:choice_of_det}
We use the function $\det(\cdot)^{1/p}$ instead of the largest eigenvalue for multiple reasons. First, note that
\[
\det( S^{-1} \hat{T}_L)^{1/p} = \left( \dfrac{\det(\hat{T}_L)}{\det(S)}  \right)^{1/p}\,. 
\]
Since the determinant of a covariance matrix of a random variable is referred to as the generalized variance of the random variable \citep{wilks:1932}, this ratio of generalized variances is akin to the ratio of variances in the univariate case. Second, the $p$th root of the determinant is the geometric mean of the eigenvalues of the matrix. Thus, the determinant accounts for variability in \textit{all} directions and not only in the direction of the principal eigenvector. The power $1/p$  ensures stability and invariance to change of units \citep{sengupta:1987}. Also, when $p=1$, \eqref{eq:vk_msprf} is the univariate PSRF in \eqref{eq:vk_psrf}. 
\end{remark}
% \begin{remark}\label{rem:pis1_univariate}
% When $p=1$, \eqref{eq:vk_msprf} is the univariate PSRF in \eqref{eq:vk_psrf}. 
% \end{remark}
\begin{remark}\label{rem:one_chain}
Users commonly run  a single Markov chain in their analysis ($m = 1$). The use of the replicated lugsail batch means estimators to estimate $T_{\infty}$ or $\tau^2_{\infty}$ allows a direct application of the GR statistic to a single chain.
\end{remark}

% subsection  (end)

% section multivariate_potential_scale_reduction_factor (end)

\section{Relation to ESS and choosing $\delta$} % (fold)
\label{sec:choosing_delta}

A challenge in implementing the GR diagnostic is choosing the PSRF cutoff, $\delta$. \cite{gel:car:2004}, say
\begin{quote}
The condition of $\hat{R}$ near 1 depends on the problem at hand; for most examples, values below 1.1 are acceptable, but  for a final analysis in a critical problem, a higher level of precision may be required.		
\end{quote}
In this section we establish and highlight the relationship between ESS and PSRF. Using the quantitative guidelines established in the literature for terminating simulation using ESS, we obtain interpretable values of $\delta$.

For an estimator, ESS is the number of independent samples with the same standard error as a correlated sample. Recall that $T_{\infty}$ is the covariance matrix in the Markov chain CLT and $\Sigma$ is the covariance matrix of the target distribution. If $p = 1$,  both $T_{\infty}$ and $\Sigma$ are scalars. For $m \geq 1$  chains, each of length 	$n$, \cite{vats:fleg:jones:2019} define ESS as
\begin{equation*}
\label{eq:ESS_def}
	\text{ESS} = mn \left(\dfrac{\det(\Sigma)}{\det(T_{\infty})} \right)^{1/p}\,.
\end{equation*}
For $p = 1$, this reduces to the following univariate definition of ESS as discussed by \cite{gong:fleg:2016} and \cite{kass:carlin:gelman:neal:1998}:
\[\text{ESS}_{p=1} = mn \dfrac{\sigma^2}{\tau^2_{\infty}}\,. \]
Strongly consistent estimators of $T_{\infty}$ and $\Sigma$ will yield a strongly consistent estimator of ESS. Thus, an estimator of ESS is the following:
\[\widehat{\text{ESS}} = mn \left(\dfrac{\det(\hat{S})}{\det(\hat{T}_L)} \right)^{1/p}\,\,. \]

% \[\widehat{\text{ESS}} = \left\{ mn \dfrac{s^2}{\hat{\tau}^2},\,\, n \dfrac{s^2}{\hat{\tau}^2}, \,\,mn \left(\dfrac{\det(\hat{\Sigma})}{\det(\hat{T})} \right)^{1/p},\,\,n \left(\dfrac{\det(\hat{\Sigma})}{\det(\hat{T})} \right)^{1/p}  \right\}\,. \]

%\cite{gong:fleg:2016} for the univariate case, and \cite{vats:fleg:jones:2019} for the general multivariate problem, provide a theoretically-justified lower bound on the number of effective samples required to obtain a certain level of precision. 
A theoretically-justified lower bound on the number of effective samples required to obtain a certain level of precision has been determined for the univariate case \citep{gong:fleg:2016} and for the general multivariate problem \citep{vats:fleg:jones:2019}.
Just as we can calculate the sample size necessary to construct a confidence interval with a desired width, we can obtain a lower bound on the ESS necessary to construct a confidence region with a desired relative volume.  %% DOOTS is this wording ok?
%This lower bound is obtained using an argument similar to a sample size calculation for constructing a confidence interval with a desired width. 
%Doots - Cool.
Suppose the goal is to make $100(1 - \alpha)\%$ confidence regions for $\mu$, using estimator $\hat{\mu}$. Let $\epsilon$ be the desired volume of the confidence region for $\hat{\mu}$ relative to the generalized standard deviation in the target distribution, $\det(\Sigma)^{1/2p}$. Then $\epsilon$---the relative volume of the confidence region---is akin to the  width of a confidence interval in sample size calculations.

Let $\chi^2_{1-\alpha, p}$ be the $(1-\alpha)$th quantile of the $\chi^2$ distribution with $p$ degrees of freedom. \cite{vats:fleg:jones:2019} show that if simulation is terminated when the estimated ESS satisfies
\begin{equation}
\label{eq:ESS_bound}
	\widehat{\text{ESS}} \geq \dfrac{(2)^{2/p} \pi}{(p \Gamma(p/2))^{2/p}} \dfrac{\chi^2_{1- \alpha, p}}{\epsilon^2}:= M_{\alpha, \epsilon,p}\,,
\end{equation}
then the confidence regions created at termination will asymptotically have the correct coverage probability. The lower bound $M_{\alpha, \epsilon, p}$ can be calculated \textit{a priori}, and simulation can terminate when the estimated ESS exceeds $M_{\alpha, \epsilon,p}$. 

It is straightforward to see that 
\begin{align*}
	\hat{R}^p_L &= \sqrt{ \left(\dfrac{n-1}{n} \right) + \dfrac{m}{\widehat{\text{ESS}}}}\\ 
	& \approx \sqrt{ 1 + \dfrac{m}{\widehat{\text{ESS}}}}\\
	& \leq \sqrt{ 1 + \dfrac{m}{M_{\alpha, \epsilon,p}}} :=  \delta_{\epsilon}\,. \numberthis \label{eq:delta}
\end{align*}
% and we define
% \begin{equation}
% \label{eq:delta}
% \delta_{\epsilon} := \sqrt{ 1 + \dfrac{m}{M_{\alpha, \epsilon,p}}} \,.
% \end{equation}
Because $\delta_{\epsilon}$ can be calculated \textit{a priori}, simulation can terminate when the PSRF drops below the threshold $\delta_{\epsilon}$. The value of $M_{\alpha, \epsilon, p}$ is obtained from \eqref{eq:ESS_bound} and is most affected by the choice of $\epsilon$ \citep{vats:fleg:jones:2019}. Therefore, the desired $\delta_\epsilon$ will be most affected by the choice of $\epsilon$. Because $\epsilon$ is interpretable,  $\delta_{\epsilon}$ is  interpretable; terminating when $\hat{R}^p_L \leq  \delta_{\epsilon}$ is equivalent to terminating simulations when $\widehat{\text{ESS}} \geq M_{\alpha, \epsilon, p}$ for  estimating the mean of the target distribution. 

\begin{example}
In our examples, we choose $\epsilon = .10$ and $\alpha = .05$. That is, for creating 95\% confidence regions, we desire the volume of the confidence region for the Monte Carlo estimator of the mean to be less than 10\% of  $\det(\Sigma)^{1/2p}$. For problems with $m = 3$ and  $p = 1$, $M_{.05, .10, 1} = 1537$, which corresponds to $ \delta_{\epsilon} = 1.000976$. Thus, the desired termination threshold in this situation is dramatically lower than the ad-hoc cutoff of 1.1. 
\end{example}

\begin{remark}
	\label{rem:minimum_effort}
\cite{vats:fleg:jones:2019} explain that a minimum simulation effort must be set to safeguard from premature termination due to early bad estimates of $\sigma^2$. We concur and  suggest  a minimum simulation effort of $n = M_{\alpha, \epsilon,p}$.
\end{remark}

% section choosing_ (end)

% section introduction (end)

\section{Examples} % (fold)
\label{sec:example}
 
%  \textbf{Simulation study plan}

%  For all the examples, we want to compare the performance of the four Gelman-Rubin-Brooks diagnostics:
%  \begin{itemize}
%  	\item Univariate $m$ chains: compare ours and traditional GR.

%  	\item Univariate single chain: compare our and GR from split chain.

%  	\item Multivariate $m$ chains: compare ours and GRB traditional method

%  	\item Multivariate single chain: compare ours and GRB split chain.
%  \end{itemize}

% In each example, we observe the following quantities:

%  \begin{enumerate}
%  	\item For examples where the true value of $\tau^2$ and $\sigma^2$ is known, compare the $\hat{\tau}^2$ and $B$ with the true $\tau^2$.

%  	\item Set $\epsilon = .10, .05, .02$ and calculate $\tilde{\delta}_{\alpha, \epsilon}$ for each of the four methods and determine when the diagnostics ask to converge. Most importantly, over replications determine how variable their GR diagnostic is and how variable our GR diagnostic is.

%  \end{enumerate}

\subsection{$t_5$-distribution continued} % (fold)
\label{sub:t_example}

\begin{figure}[htb]
	\centering
\includegraphics[width=2.2in]{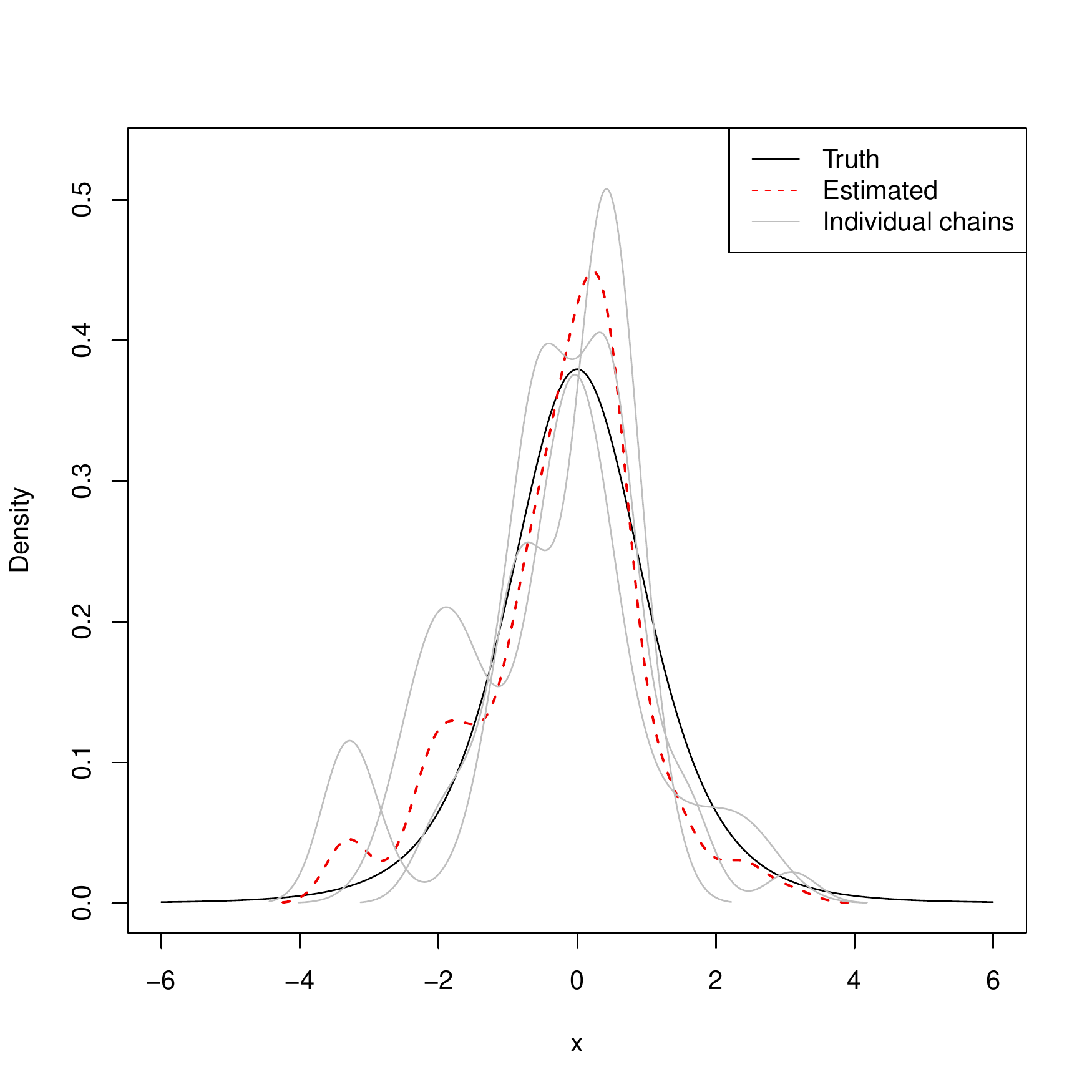}
 \includegraphics[width=2.2in]{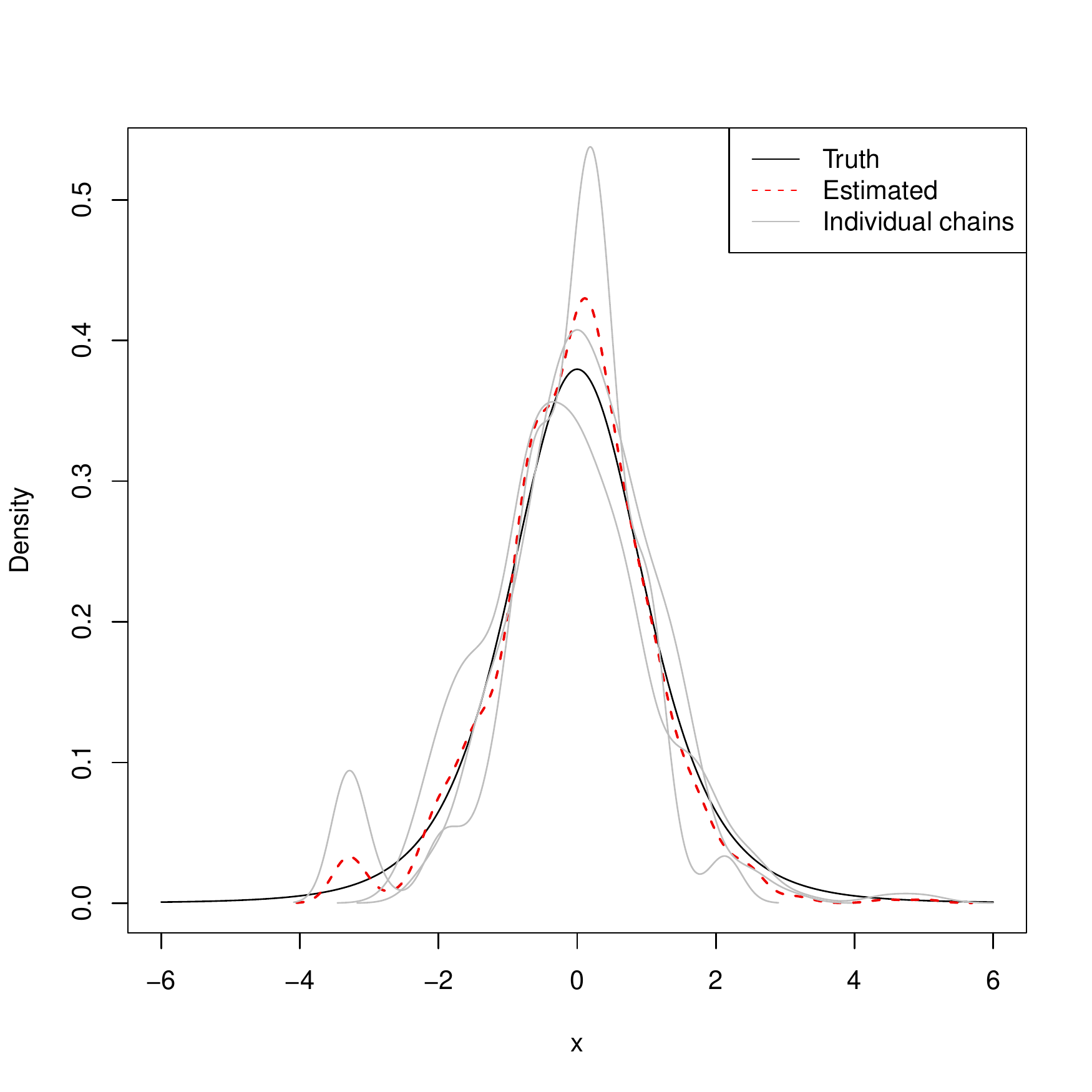}
 \includegraphics[width=2.2in]{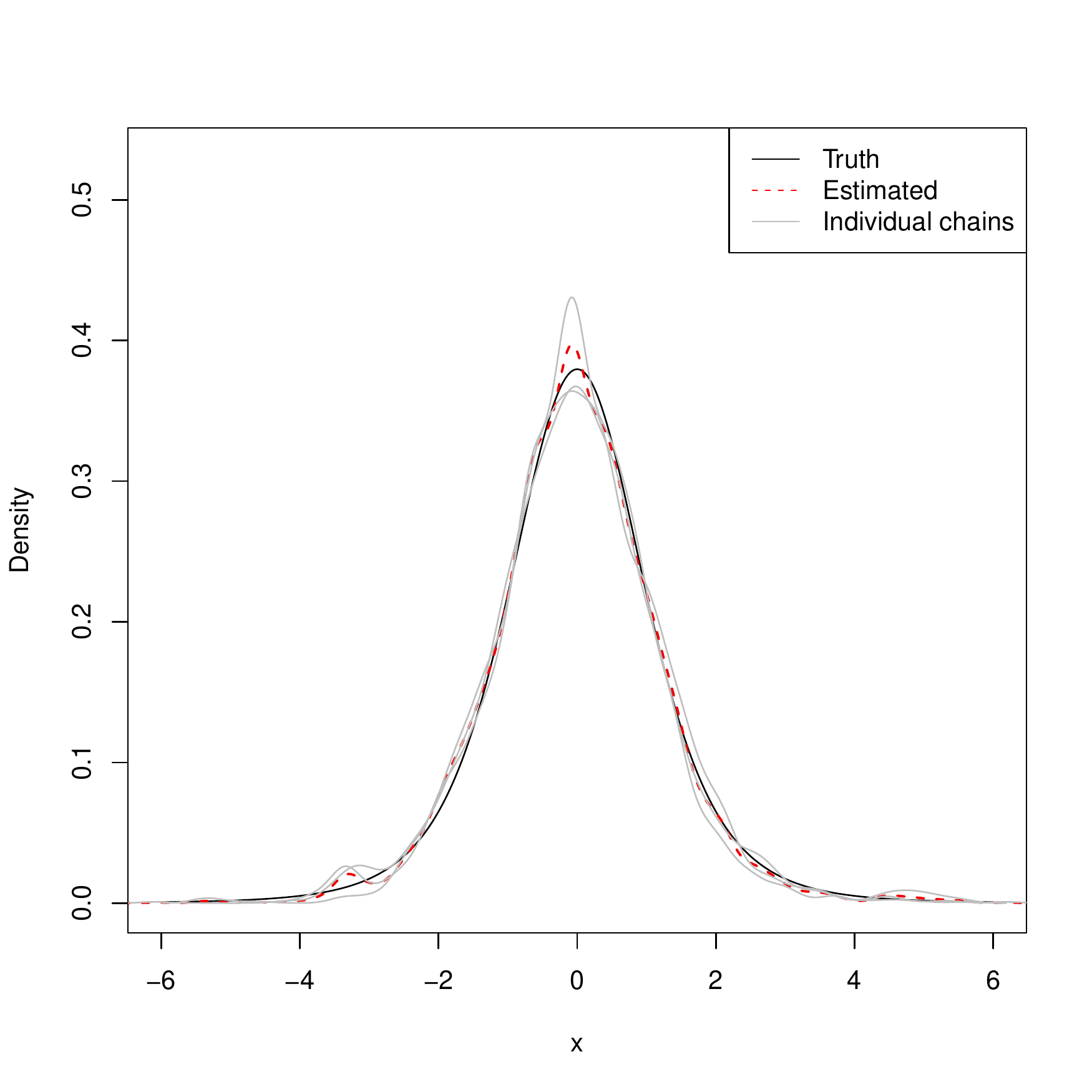}
 \includegraphics[width=2.2in]{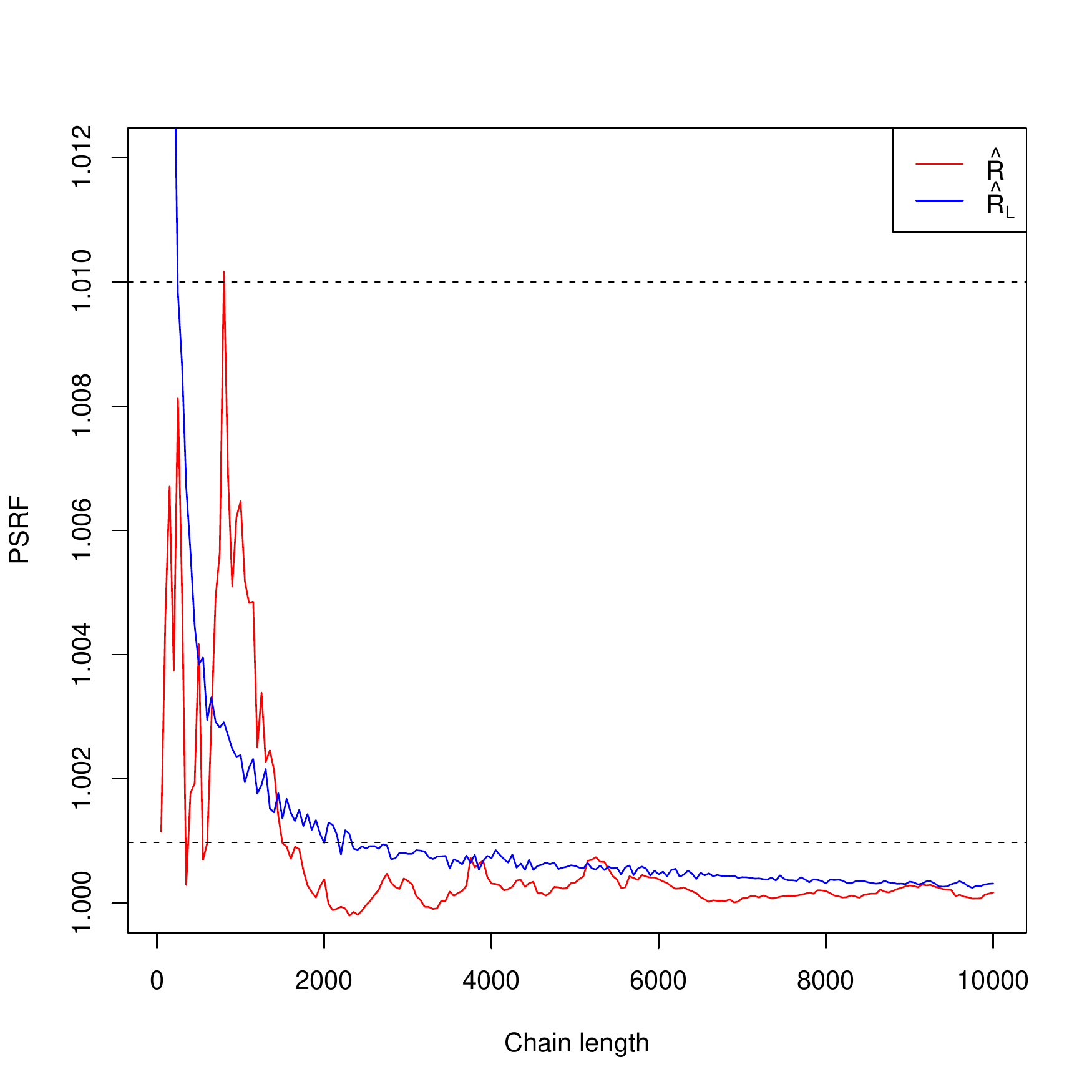}
	\caption{$t$-distribution: (From left to right and top to bottom). Estimated density using $\hat{R}_L$ with  $\delta = 1.1$, $\delta = 1.01$ and $\delta_{.1} = 1.000975$. Bottom right is a running plot of the estimated PSRF using both $\hat{R}$ (red) and $\hat{R}_L$ (blue). The horizontal dotted black line is $\delta_{.1}=1.000975$.}
	\label{fig:t_complete}
\end{figure}

Recall the  $t_5$-distribution example introduced in Section \ref{sec:introduction}, where we run $m = 3$ chains with starting values randomly sampled from a $t_2$-distribution. For our seed, the starting points were $(0.484, 1.370, -0.131)$. For $\delta = 1.1, \delta= 1.01$, and $\delta_{.1} = 1.000975$, we check whether each convergence criterion is satisfied for $\hat{R}_L$ in increments of 50 iterations and present the estimated density plots in Figure~\ref{fig:t_complete}. For $\delta = 1.1$ and $\delta = 1.01$, the termination criteria are met at $n = 50$. The density estimate clearly indicates poor quality of estimation. For $\delta_{.1} = 1.000975$, the sampler terminates at $n = 2350$ iterations, resulting in improved estimation. Further smaller values of $\epsilon$ will provide further improvements, and $\epsilon$ can be chosen based on the quality of estimation desired. In Figure~\ref{fig:t_complete}, we also present a running plot of $\hat{R}$ and $\hat{R}_L$ which illustrates the erratic behavior of $\hat{R}$, especially for small sample sizes. In comparison, $\hat{R}_L$ is far more stable and exhibits monotonic decreasing behavior (up to randomness). 
% subsection subsection_name (end)

 \subsection{Autoregressive process of order 1} % (fold)
 \label{sub:ar}

Consider the autoregressive process of order 1 (AR(1)). For $t = 1, 2, \dots$, let $Y_t \in \real$ and $\epsilon_t \sim N(0, \nu^2)$. For $|\rho| < 1$, the AR(1) process is
\begin{equation*}
	Y_t = \rho Y_{t-1} + \epsilon_{t}\,.
\end{equation*}
This describes  a Markov chain with stationary distribution $N(0, \sigma^2)$, where 
\begin{equation}
	\label{eq:sigma2_ar1}
	\sigma^2 = \dfrac{\nu^2}{1 - \rho^2}\,.
\end{equation}
The autocorrelation coefficient $\rho$ determines the rate of convergence of the Markov chain. In particular, if $|\rho| < 1$  a Markov chain CLT holds for $\bar{Y}_n = n^{-1} \sum_{t=1}^{n} Y_t$ with the following asymptotic variance:
 \begin{equation*}
 	\tau^2_{\infty} = \sigma^2 \; \dfrac{1 + \rho}{1 - \rho}\,.
 \end{equation*}
For finite $n$, we can obtain an expression for $\tau^2_n = n\Var_F(\bar{Y}_n)$,
\begin{align*}
\tau^2_n 
% &=  \Var(Y_{1}) + 2 \sum_{k=1}^{n-1} \left( \dfrac{n-k}{n}  \right)\text{Cov}(Y_{1}, Y_{(1+k)}) \\ 
& = \sigma^2 + 2 \sigma^2\sum_{k=1}^{n-1} \left( \dfrac{n-k}{n}  \right)\rho^k\,. \numberthis \label{eq:ar1_true_tau_n}
\end{align*}

\begin{figure}[htb]
	\centering
\includegraphics[width=2.5in]{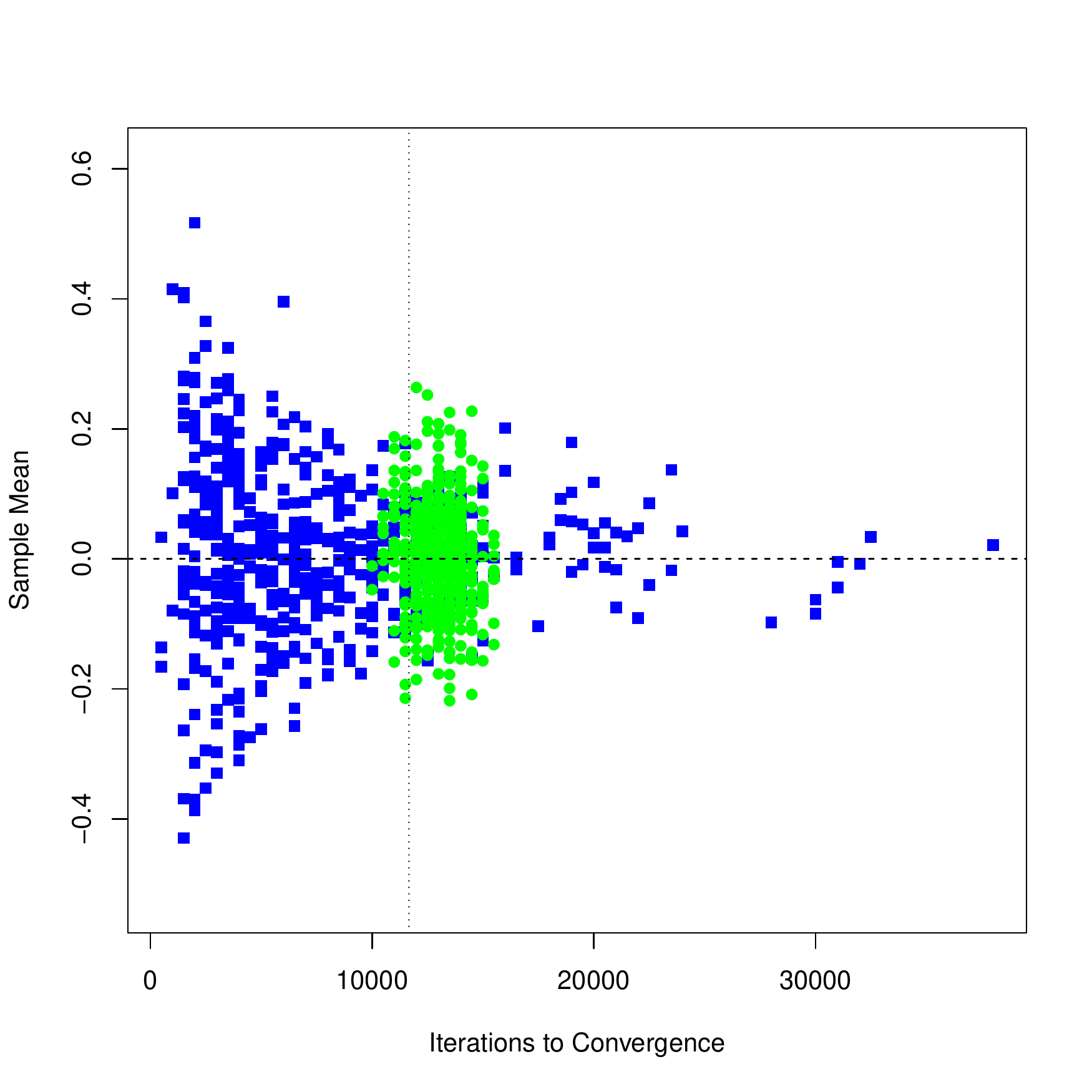}
 \includegraphics[width=2.5in]{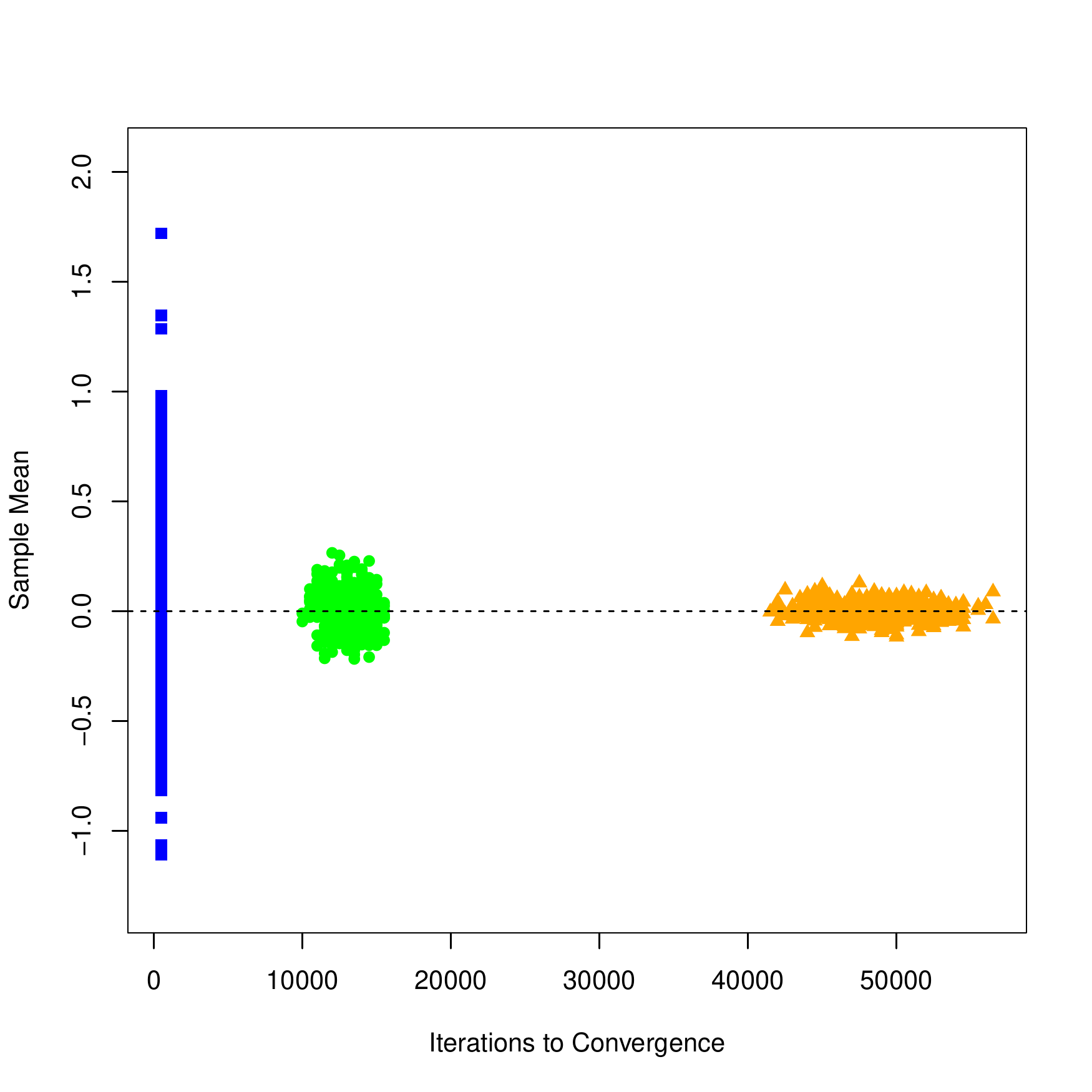}
	\caption{AR(1): Sample mean versus termination iteration for $m = 5$ chains. (Left) Blue squares are for $\hat{R}$; green points are for $\hat{R}_L$. The horizontal dotted line is the true mean. The vertical dotted line is the true chain length for termination cutoff $\delta_{.10}$. (Right) Blue squares are for $\hat{R}_L \leq \delta = 1.1$. Green points and orange triangles are for $\hat{R}_L \leq\delta_{.10}$ and $\hat{R}_L \leq\delta_{.05}$, respectively. The horizontal dotted line is the true mean.}
	\label{fig:ar1_cloud_test_lugsail}
\end{figure}

In this example, we set $\nu = 1$ and $\rho = .95$. Since the true values of $\tau^2_n$ and $\sigma^2$ are known, we can compare the performance of our proposed methods with that of the original GR methods. Over 500 replications, we determine when $\hat{R}$ and $\hat{R}_L$ reach $ \delta$ and record the Monte Carlo estimate, $\bar{Y}_n$, and termination sample size; each criterion is checked in increments of 500 iterations. In Figure~\ref{fig:ar1_cloud_test_lugsail}, we plot $\bar{Y}_n$ at termination versus the termination index using both $\hat{R}$ and $\hat{R}_L$ for a single chain and for $m = 5$ chains. For these simulations, we set $\epsilon = {.10}$  and use $m = 5$; this  yields termination threshold $\delta_\epsilon = 1.001625$. We compare our results against the true value of the PSRF determined by \eqref{eq:sigma2_ar1} and \eqref{eq:ar1_true_tau_n}. 

% Similar conclusions can be drawn from the $m=5$ and $m=1$ settings (and their corresponding plots in Figure \ref{fig:ar1_cloud_test_lugsail}).
In Figure~\ref{fig:ar1_cloud_test_lugsail}, we present our simulation results. First, we inspect the horizontal variability by  comparing the  number of iterations required for convergence for the two convergence statistics. The variability in the termination procedure using $\hat{R}$ is large: some runs converged almost immediately while others required over 30,000 steps. The replicated lugsail batch means estimators  terminate close to the true termination index and do so with considerably lower variability; this follows from the efficiency result in \eqref{eq:efficiency_comparison}.   Second, we inspect the vertical variability in Figure \ref{fig:ar1_cloud_test_lugsail}: the means produced at termination by $\hat{R}_L$ have low, near-uniform variability in each plot while the original GR diagnostic produces means with more variability.

\begin{figure}[htb]
	\centering
 \includegraphics[width=2.5in]{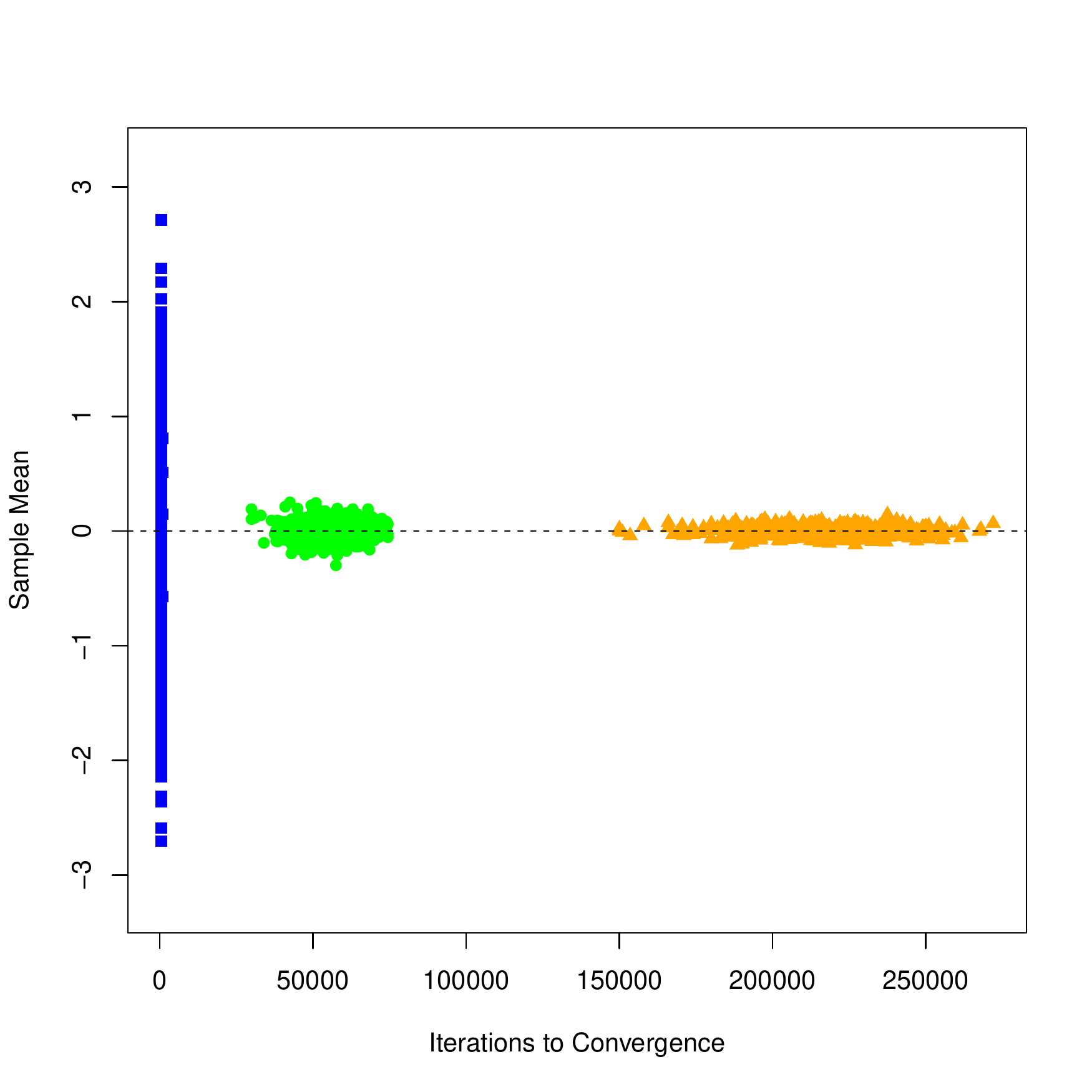}
	\caption{AR(1): Sample mean versus termination iteration for $m = 1$ chain. Blue squares are for $\hat{R}_L \leq \delta = 1.1$. Green points and orange triangles are for $\hat{R}_L \leq\delta_{.10}$ and $\hat{R}_L \leq \delta_{.05}$, respectively. The horizontal dotted line is the true mean.}
	\label{fig:ar1_cloud_test_lugsail_1chain}
\end{figure}

Unlike $\hat{R}$, $\hat{R}_L$ can be calculated for a single chain. In Figure~\ref{fig:ar1_cloud_test_lugsail_1chain}, we plot the iterations to convergence for $\hat{R}_L$ using three convergence criteria versus the estimated sample mean at convergence. The plot here is essentially similar to the right plot in Figure~\ref{fig:ar1_cloud_test_lugsail} in that it is clear that $\delta = 1.1$ yields high variability in the resulting estimates. 

 % Since the original GR diagnostic requires multiple chains, $\hat{R}$ is calculated using split-$\hat{R}$ when $m=1$; that is, the GR statistic is calculated by breaking one chain of length $n$  into $m = 2$ chains, each of length $n/2$. Split-$\hat{R}$ terminates the Markov chain much too early. This is because the termination criterion in \eqref{eq:delta} for $m = 1$ using $\delta_{.10}$ no longer holds for split-$\hat{R}$ since we have artificially broken a single chain into two chains. As a result, the process of determining a cutoff for split-$\hat{R}$ is unclear. Since lugsail estimation does not require multiple chains, $\hat{R}_L$ does not experience this problem. 
%

For three different termination criteria we  calculate the iterations to convergence using $\hat{R}_L$ and the Monte Carlo average at convergence.  Results are in the right plot of Figure~\ref{fig:ar1_cloud_test_lugsail}. Naturally, smaller values of $\epsilon$---which correspond to smaller values of $\delta_\epsilon$---yield later termination. Most importantly, we note the poor performance of the ad-hoc $\hat{R}_L \leq 1.1$ criterion: the variability in the sample mean is much too large to yield any confidence in the quality of estimation. 
% We also note that the overall simulation effort ($mn$) using $m = 1$ and $m = 5$ is roughly the same.
% \begin{figure}[htb]
% 	\centering
%  \includegraphics[width=2.5in]{612WithoutLegend5Ch.pdf}	
%  \includegraphics[width=2.5in]{612WithoutLegend1Ch.pdf}
% 	\caption{Monte Carlo average versus the number of iterations to convergence for three termination criteria for $m = 5$ chains (left) and $m = 1$ chain (right). Blue squares are for $\hat{R}_L < 1.1$. Green points and orange triangles are for $\delta_{.10}$ and $\delta_{.05}$, respectively. The horizontal dotted line represents the true mean.}
% 	\label{fig:ar1_cloud_test_delta}
% \end{figure}

 % subsection ar (end)

\subsection{Bimodal Gaussian distribution} % (fold)
\begin{figure}[htbp]
	\centering
	\includegraphics[width = 5in]{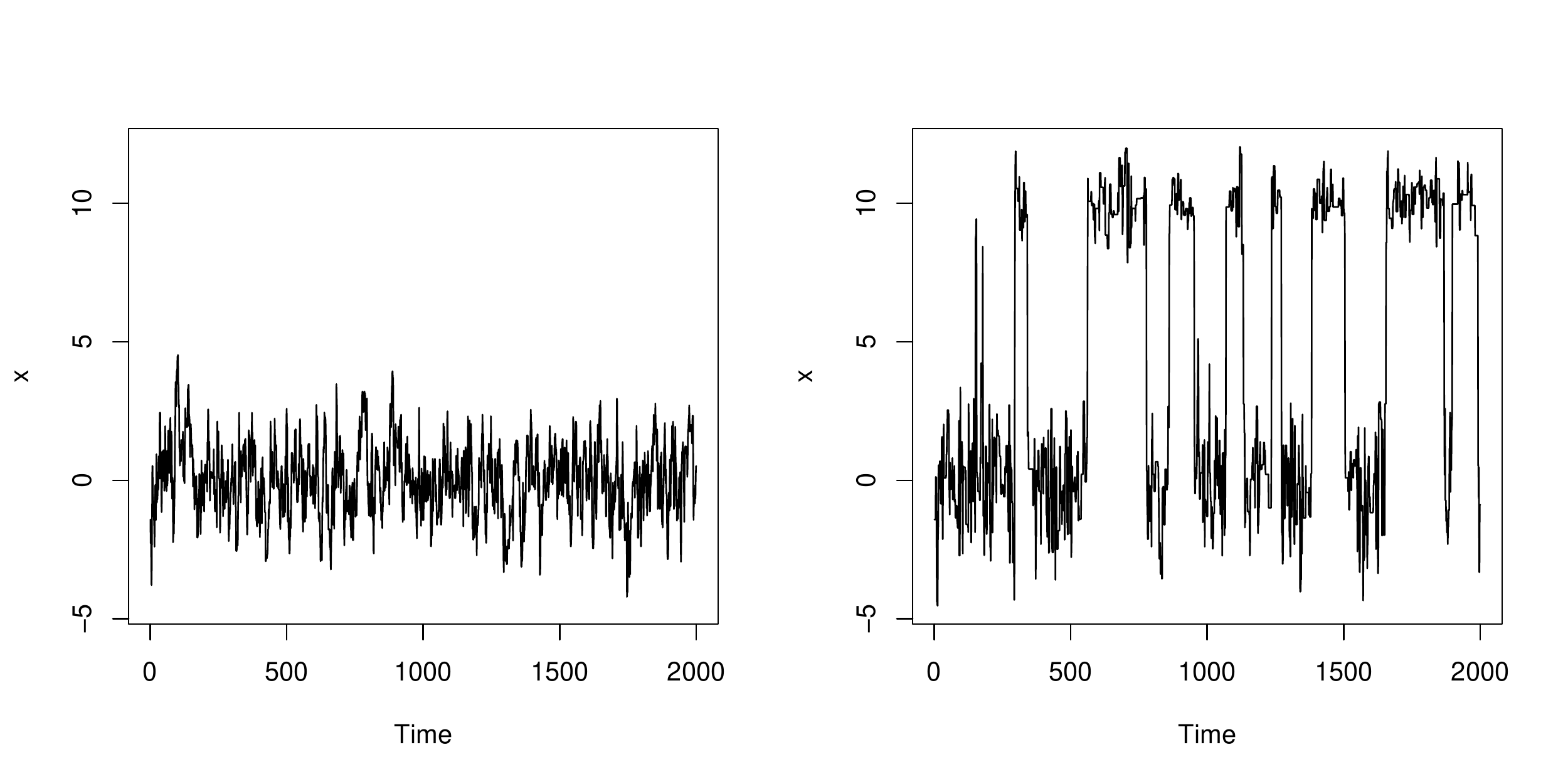}
	\caption{Bimodal: Trace plots of one Markov chain run for $h = 1$ (left) and $h = 10$ (right).}
	\label{fig:mix_trace}
\end{figure}

\label{sub:bimodal_gaussian_target}
% \begin{figure}[tb]
% 	\centering
% 	\includegraphics[width = 3in]{mixG_density.pdf}
% 	\caption{The mixture of normals target density.}
% 	\label{fig:mix_density}
% \end{figure}
Let $f(x;\theta, \lambda^2)$ be the density of a normal distribution with mean $\theta$ and variance $\lambda^2$. Consider the following density of a mixture distribution of two normal random variables:
\[
f(x) = \dfrac{1}{2} f_1(x; 0,2) + \dfrac{1}{2} f_2(x; 10, .5)\,.
\]
We run a random walk Metropolis-Hastings MCMC algorithm with proposal distribution $N(\cdot, h)$ and consider two choices of $h$:  $h = 1$ and  $h = 10$. A larger $h$ allows the Markov chains to jump between modes relatively easily so that each Markov chain explores the state space relatively well. The first setting with $h = 1$ localizes each Markov chain, not allowing them to easily jump modes. Trace plots illustrating this behavior are in Figure~\ref{fig:mix_trace}.

\begin{figure}[!htb]
	\centering
	\includegraphics[width = 2.4in]{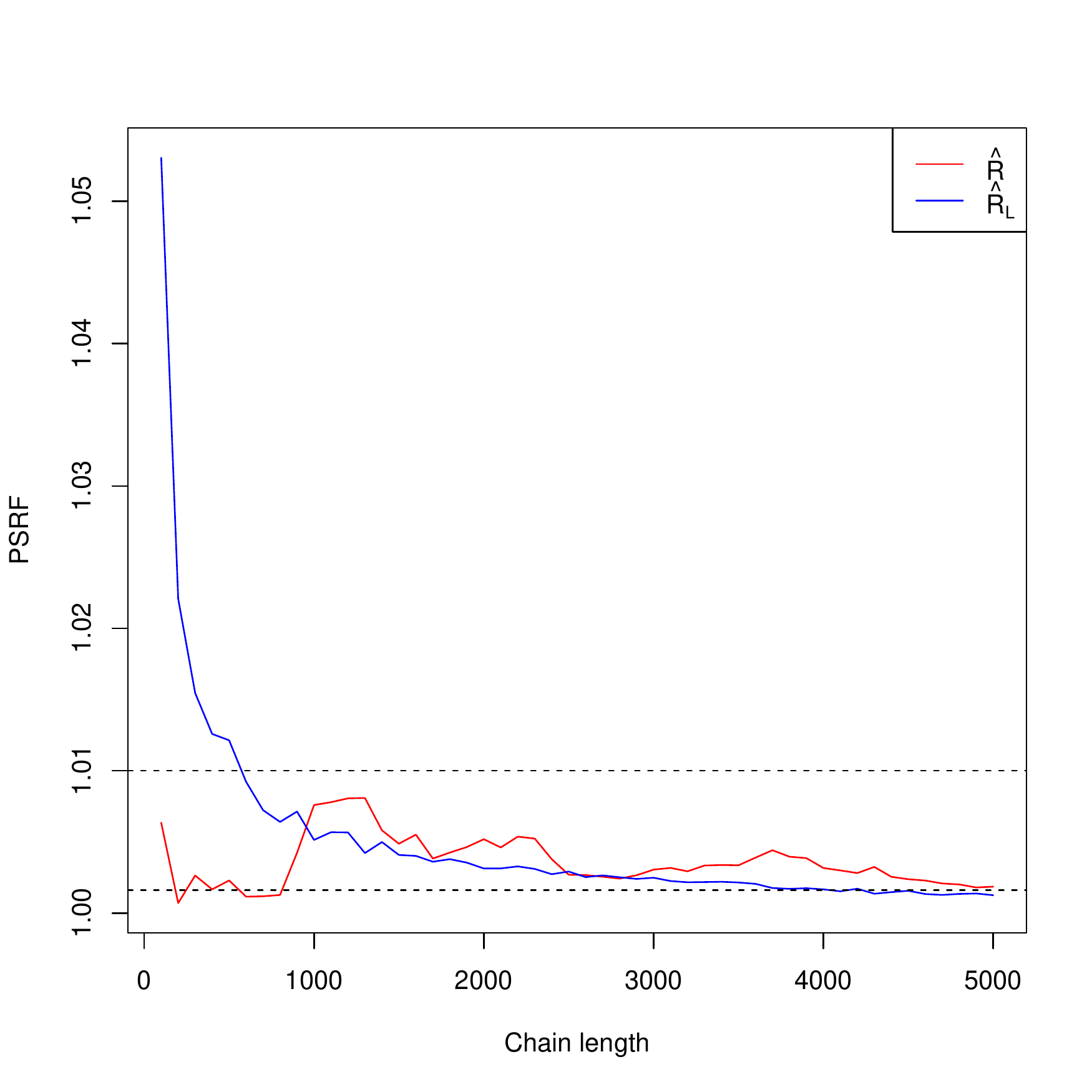}
	\includegraphics[width = 2.4in]{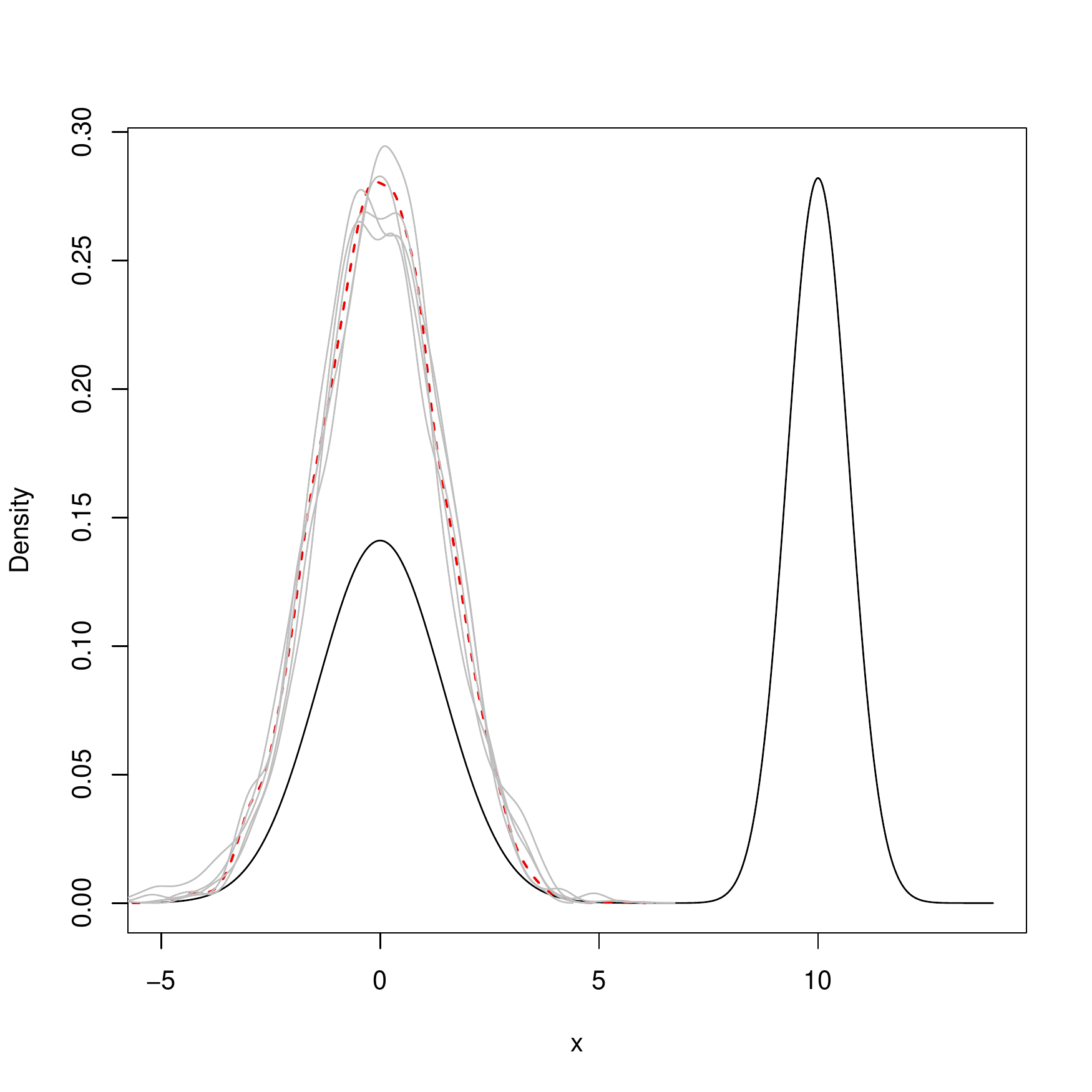}
	\caption{Bimodal: Running plot of PSRF estimates  with horizontal lines drawn at $\delta = 1.01$ and $\delta _{.1} = 1.001625$ (left) and estimated density plot at termination using $\hat{R}_L \leq\delta_{.1}$ (right).}
	\label{fig:mixbad_early}
\end{figure}

For $h = 1$, we run $m = 5$ Markov chains from the first mode and track both $\hat{R}$ and $\hat{R}_L$. Since the Markov chains do not adequately explore the state space---in particular, the chains have not discovered the second mode yet---both methods prematurely diagnose convergence; see Figure~\ref{fig:mixbad_early} where the Markov chains satisfy $\hat{R}_L < \delta_{.1} = 1.001625$ at $n = 4100$. Since this declaration of convergence is premature, the estimated density plot at termination is nowhere near the truth. The GR diagnostic, even with our improvements, cannot possibly detect lack of convergence when the chain has failed to travel to areas of critical mass.  It is thus imperative to choose an MCMC sampler that adequately explores the state space before any output analysis is considered. 
% (Lower left) Running plot of estimated PSRF after 3e5 iterations; $\hat{R} \leq \delta_{.1}$ at 8.5e5 iterations. (Lower right) Estimated density plot using $\hat{R} \leq \delta_{.1}$ and a minimum simulation effort of 8.5e5 iterations.
% DOOTS  the text says "declares convergence according to $\delta_{.1}$ at chain length 8.5e5" so I am changing 8.7e5 to that so be consistent but please check to see which is correct (and then correct in both places)
% DOOTS another thing: Lower left used to say "Running plot of estimated PSRF after 8.5e5 iterations" and I changed 8.5e5 to 3e5
% DOOTS another thing: for lower right, do we really mean Rhat or do we mean Rhat_L
% DOOTS do you 
% here's what LOWER RIGHT used to say:
% Estimated density plot for when $\hat{R} \leq \delta_{.1}$ after 3e5 iterations. }

\begin{figure}[htb]
	\centering
	\includegraphics[width = 2.5in]{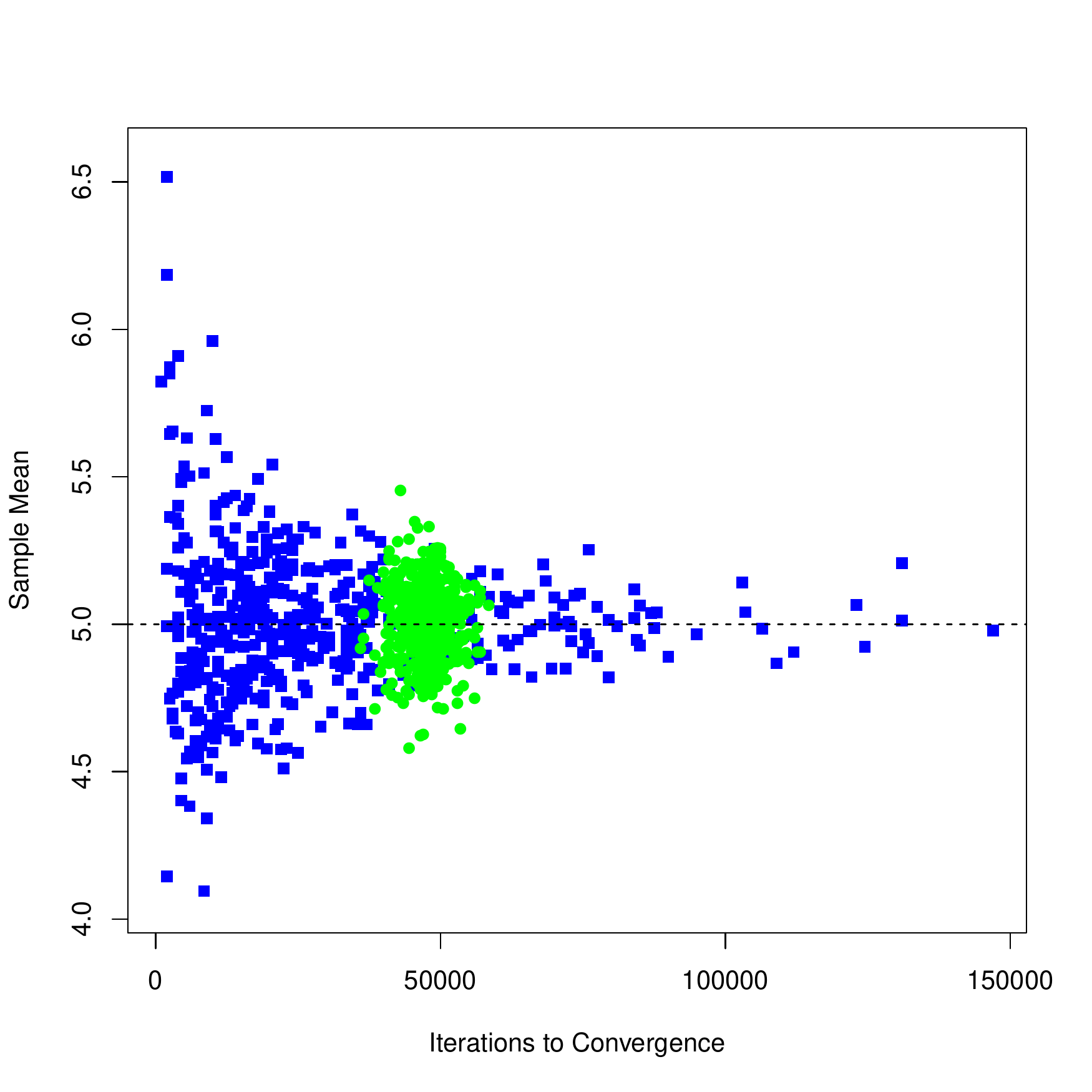}
		\includegraphics[width = 2.5in]{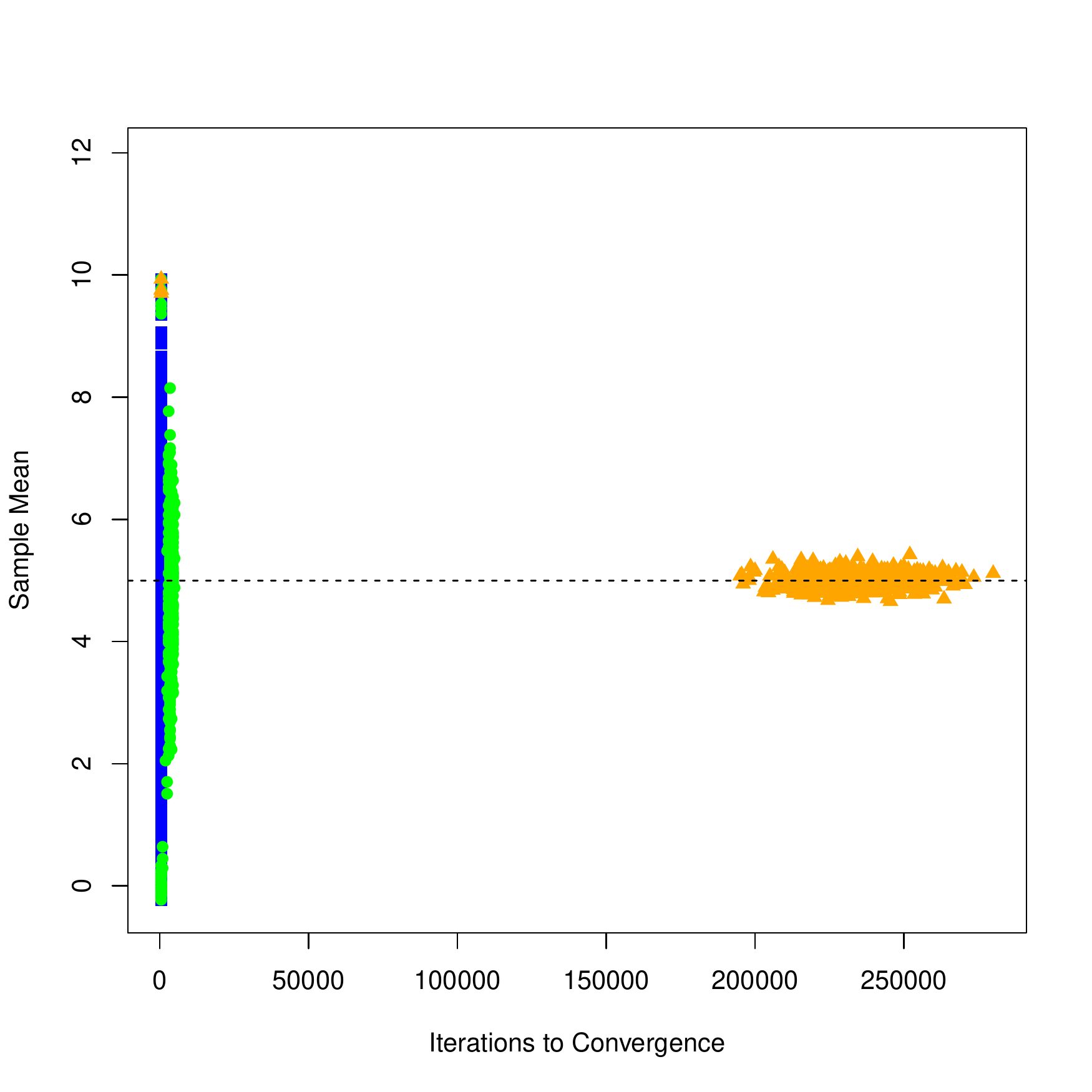}
	\caption{Bimodal: Sample mean versus Monte Carlo sample size at termination with the horizontal dotted line being the true mean. (Left) For $m = 5$ chains with termination threshold $\delta_{.1}$. Blue squares are obtained using $\hat{R}$; green points are obtained using $\hat{R}_L$. (Right) For $m = 1$ chain with statistic $\hat{R}_L$ and three termination thresholds. Blue squares are obtained using $\delta = 1.1$, green points are obtained using $\delta = 1.01$, and orange triangles are obtained using $\delta_{.1}$.}
	\label{fig:mixgood_cloud}
\end{figure}

For $h = 10$, the Markov chain is able to move across modes often so that sample quantities are well estimated. Over 500 replications, we run $m = 5$ Markov chains starting from an over-dispersed distribution. In each replication, we record  the chain length at  $\hat{R} \leq \delta_{.1}$ and $\hat{R}_L \leq \delta_{.1}$. Results are presented in the left plot of Figure~\ref{fig:mixgood_cloud}. Using $\hat{R}$ results in termination as early as chain length 500 and as late as chain length 1.5e5. In contrast,  $\hat{R}_L$ has far less variability in the termination time and in the sample mean estimates at termination.
% This corresponds to fairly constant variability in the sample means associated with chains that have been terminated based on statistic $\hat{R}_L$. 
%DOOTS is this wording ok? - FURHER REWORDED.
%In addition, the  variability in the resulting sample mean is fairly constant across the termination chain lengths with $\hat{R}_L$. 

In order to assess the performance of $\hat{R}_L$ for a single chain, we implemented another simulation study using only $\hat{R}_L$ for $m = 1$ chain with cutoffs $\delta = 1.1$, $\delta = 1.01$, and $\delta_{.1} = 1.000325$; the results are presented in the right plot of Figure~\ref{fig:mixgood_cloud}. % DOOTS I changed this ref to fig:mixgood_cloud but fix it if it should indeed be fig:ar1_cloud_test_lugsail - WELL SPOTTED
Barring two of the 500 replications where the single chain was not able to jump from the local mode,  termination criterion $\delta_{.1}$ dramatically stabilizes the estimation quality at termination. The  convergence criteria of $\delta = 1.1$ and $\delta = 1.01$ lead to premature termination.

% {\color{red} add summarizing paragraph to drive home that rhat L works better whether h=1 or h=10}

% subsection bimodal_gaussian_target (end)

 \subsection{Bayesian logistic regression: Titanic data} % (fold)
 \label{sub:Titanic}

On April 15, 1912, the RMS Titanic sank after colliding with an iceberg on its maiden voyage. The accident killed 1502 of the 2224 passengers and crew on board.  The \texttt{titanic\_train} data in the R package \texttt{titanic}  contains information on 891 passengers aboard the Titanic and whether they survived the tragedy or not. Additional information includes the class of the passenger (Pclass, a factor with three levels), sex (a factor with two levels), age, the number of siblings/spouses aboard (SibSp), the number of parents/children aboard (Parch), the passenger's fare (Fare), and port of embarkation (Embarked, a factor with three levels).  The data-set contains 179 entries with missing values, which we remove, yielding 712 observations.

We fit a Bayesian logistic regression model to this data. Let $Y_1, \dots, Y_{712}$ be the observed binary response. $Y_i = 1$ if the $i$th passenger survived and $Y_i = 0$ otherwise. For $i = 1, \dots, 712$, let $x_i = (x_{i1}, \dots, x_{i10})^T$ denote the vector of covariates for the $i$th response.  For $\beta \in \real^{10}$, the Bayesian logistic regression setup is 
\[Y_i  \mid \beta \sim \text{Bernoulli} \left(\dfrac{1}{1 + \exp(-x_i^T\beta)}  \right)\,. \]
We assume a multivariate normal prior on $\beta$ (that is, $\beta \sim N(0, \sigma^2_{\beta}I_{10})$, where $I_{10}$ is the $10 \times 10$ identity matrix). We set $\sigma^2_{\beta} = 100$ to yield a diffuse prior on $\beta$. A random walk Metropolis-Hastings sampler available in the R package \texttt{MCMCpack} is used to sample from the intractable posterior. We tune the step size of the sampler to approximate the optimal acceptance probabilities indicated by \cite{rob:gel:gilks}.

Since posterior distribution is 10-dimensional, we employ  the multivariate PSRF to determine the number of samples required. We run $m=5$ parallel chains with starting values from across -3 to 3 standard deviations from the maximum likelihood estimate of $\beta$. 
% We first set $n = 50$ and, in increments of 10\%, check whether the multivariate PSRFs are below 1.1 and $1 + \delta_{.10}$ in \eqref{eq:delta}. 
We start with $n=50$ and---as long as the multivariate PSRFs are above 1.1 and $ \delta_{.10}$ in \eqref{eq:delta}---we increase the Markov chain length by 10\%. 

\begin{figure}[htb]
	\centering
\includegraphics[width=5in]{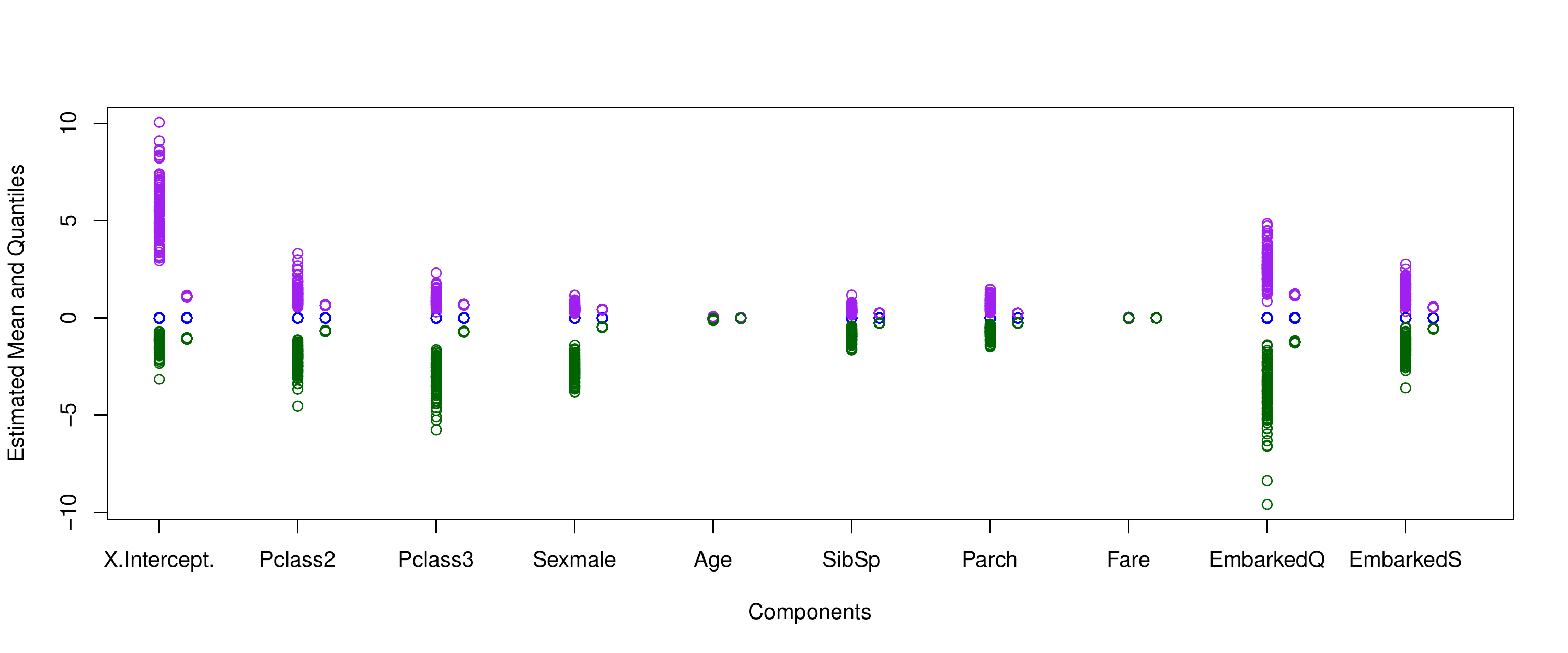}
	\caption{Titanic: Centered posterior mean and 95\% credible interval estimates from 100 replications for all 10 components. Purple circles are upper quantiles, blue circles are posterior means, and green circles are lower quantiles. Each component has two sets of points: the left points are for $\hat{R}^p_L \leq 1.1$ and the right points are for $\hat{R}^p_L \leq\delta_{.10}$}
	\label{fig:titanic_credible}
\end{figure}

In 100 replications, we note the posterior mean of $\beta$ and the 95\% credible interval at termination using both criteria. The results are in Figure~\ref{fig:titanic_credible}. It is immediately clear that the ad-hoc threshold of $\delta =  1.1$  yields  credible intervals with unacceptably large variability, as illustrated by the left set of points in Figure~\ref{fig:titanic_credible};  in this example,   the $\delta =  1.1$ cutoff  yields untrustworthy estimates.   
 In contrast, $\delta_{.10}$ produces credible interval estimates with minimal variability. 

\begin{figure}[htb]
	\centering
\includegraphics[width=3in]{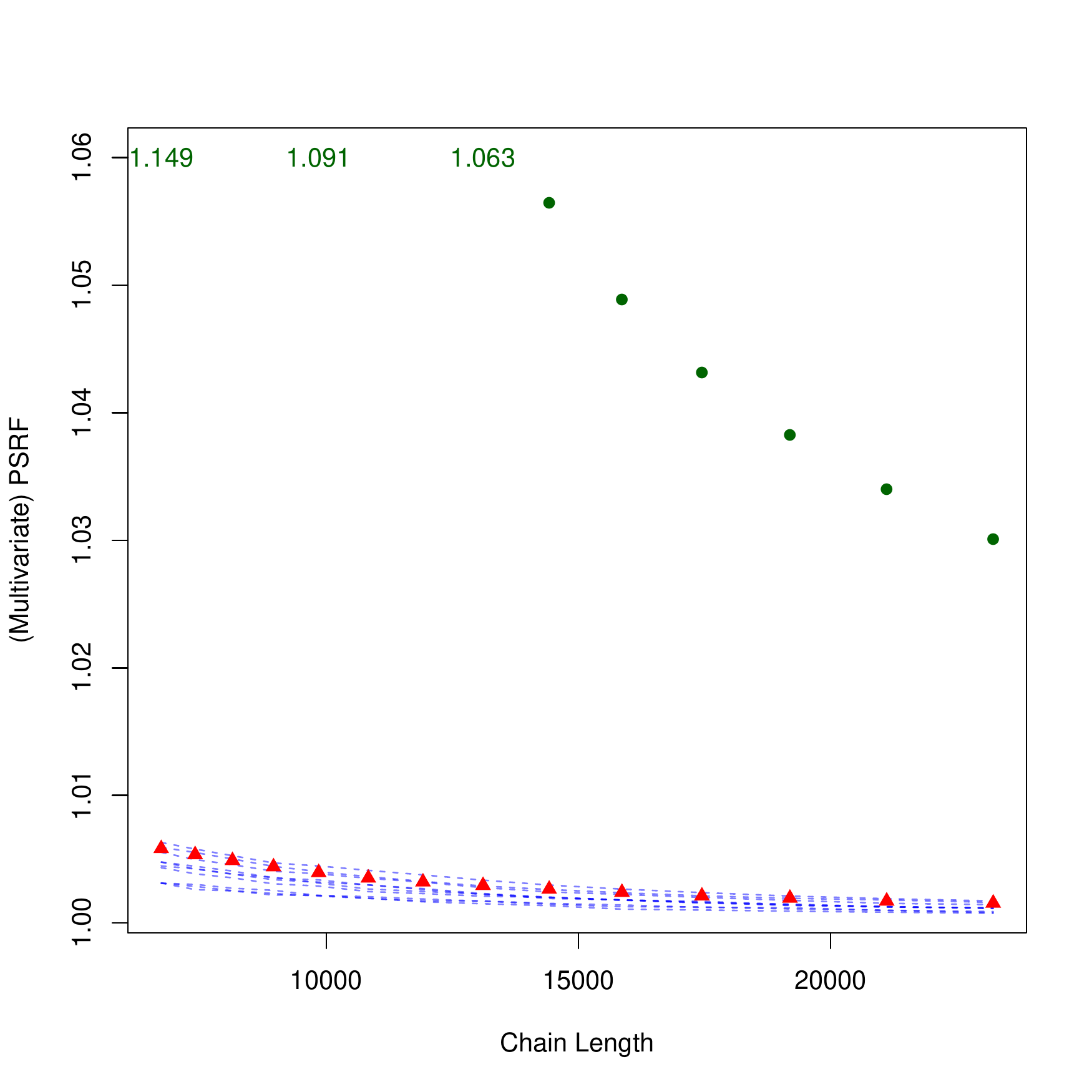}
	\caption{Titanic: Univariate PSRFs (blue dotted lines), multivariate PSRFs using the largest eigenvalue (green circles), and multivariate PSRFs using the determinant (red triangles) from $m = 5$ parallel chains. Green text represents the PSRFs of circles that did not fit in the graph.}
	\label{fig:titanic_compare_mpsrf}
\end{figure}

Next we compare the performance of the multivariate PSRF using the determinant against the performance of the original multivariate PSRF in \eqref{eq:original_mpsrf}, which uses the  largest eigenvalue \citep{bro:gel:1998}. In Figure~\ref{fig:titanic_compare_mpsrf}, we track the evolution of the two statistics, along with the 10 univariate PSRFs, for one run of the 5 parallel chains. The determinant PSRF yields values close to the univariate PSRFs, but the largest eigenvalue PSRFs are markedly more conservative, resulting in delayed termination. If conservative termination is desirable, we recommend adhering to the determinant-based multivariate PSRF and using a smaller $\epsilon$ in order to retain the ESS interpretation of the procedure. 

% Out of interest, in each of the 100 replications, we also calculated the estimated probability of survival for the lead characters in the 1997 feature film, ``Titanic''; the names of the characters are Jack and Rose, and their covariates are set using details presented in the movie. Boxplot of estimated probabilities of survival is presented in Figure~\ref{fig:jack_rose}. Unsurprisingly, the estimated probability of survival for Jack is fairly low, and probability of survival for Rose is close 1. It is important to note the significantly higher variability in estimates arising from $\delta = .1$ criterion.

% \begin{figure}[htbp]
% 	\centering
% \includegraphics[width=3in]{./../CleanSims/BayesLogistic/output/plots/SurvivalMedianTicketPriceProb.pdf}
% 	\caption{Posterior predictive probabilities}
% 	\label{fig:jack_rose}
% \end{figure}

% section example (end)

\section{Discussion} % (fold)
\label{sec:discussion}

The MCMC community has  long held the view that the GR diagnostic is susceptible to  premature and unreliable convergence diagnoses \citep{fleg:hara:jone:2008}. This certainly remains true in situations where the Markov chains are all stuck in a local mode (as demonstrated in Section~\ref{sub:bimodal_gaussian_target}). 
In this situation, we demonstrate the poor performance of $\delta=1.1$---as \cite{veht:gelman:sim:2019} also acknowledged---and emphasize the importance of choosing a well-informed termination threshold.
%In this situation, we demonstrate why it is crucial to choose a well informed cut-off and $\delta = 1.1$ does not work well, as also pointed out by \cite{veht:gelman:sim:2019}. 
Even if the Markov chains explore different parts of the state space,  we propose changes that strengthen the GR diagnostic in two significant ways: 1) we stabilize the GR statistic using improved estimators of Monte Carlo variance and 2) we safeguard against premature diagnosis by replacing $\delta = 1.1$ with  a principled, ESS-based termination threshold. Our diagnostic is 	available for public use in the \texttt{R} package \texttt{stableGR} \citep{knud:vats:2020}.

To stabilize the GR statistic, we incorporate an efficient estimator of the variance of the Monte Carlo average: the replicated lugsail batch means estimator. Our examples demonstrate how this incorporation effectively stabilizes the time-to-convergence and the resulting sample means.
An immediate advantage of the replicated lugsail batch means estimator is it can be calculated for a single chain; single chain output analysis has long been part of MCMC practice and our proposed GR statistic can easily assess convergence in this scenario. Ordinary batch means estimators and spectral variance estimators can also handle a single chain and might yield even higher statistical efficiency, but they do not naturally overestimate the Monte Carlo standard errors. 
This biased-from-above property of the lugsail estimator  safeguards the statistic against early termination. Although we believe that the replicated lugsail batch means estimator is currently the best candidate for the GR statistic, univariate and multivariate Monte Carlo variance estimation is a rich, ongoing area of research: the GR statistic will benefit from continual adaptation to incorporate advances in this area.

To address premature convergence diagnoses, we inspect the PSRF threshold of 1.1. and through various example show that a premature convergence diagnosis is often due to the arbitrary PSRF threshold of 1.1. We establish a one-to-one mapping between PSRF and ESS and use this to show that a PSRF termination threshold of 1.1   yields approximately 5 effective samples per chain, which is far too small for any reasonable number of chains. 
We then leverage this ESS-PSRF connection to construct a principled, ESS-based PSRF termination threshold. This connection makes PSRF thresholds interpretable and theoretically-motivated. Additionally, this ends the tension between ESS and PSRF---which have historically competed as methods for output analysis---by recognizing these methods are one and the same when interest is in estimating the mean of the target distribution. When interested in estimating the expectation of a general function $g$, $\int g F(dx)$, where $g(x) \ne x$, then ESS pertains to estimating $\E_Fg$ whereas the untransformed PSRF still connects to the effective sample size in estimating $\E_FX$. For the connection to remain, the PSRF must be calculated for the transformed process, $g(X)$.

Finally, we note that a significant amount of theoretical detail has been intentionally left undiscussed in order to focus on the more practical issues of the GR diagnostic implementation. We have assumed the existence of a  Markov chain central limit theorem, which requires mixing and moment conditions. Strong consistency and variance expressions for the replicated lugsail batch means estimators also require similar moment and mixing conditions. More details on the theoretical aspects of this work can be found in \cite{gupta:vats:2020} and \cite{jone:2004}.

% section discussion (end)

\section{Acknowledgments} % (fold)
\label{sec:acknowledgments}
The authors thank the anonymous referees for their feedback and comments which significantly improved the manuscript. The authors also thank Samuel Livingstone for a useful conversation that led to significant improvements in the paper. 
% section acknowledgments (end)

\bibliographystyle{apalike}
\bibliography{mcref}
\end{document}